\documentclass[utf8]{FrontiersinVancouver}

\setcitestyle{compress,square} % for articles in the journals "Frontiers in Physics" and "Frontiers in Applied Mathematics and Statistics" 
\usepackage[export]{adjustbox}
\usepackage{amsfonts}
\usepackage{datetime}
\usepackage{fmtcount}
\usepackage{etoolbox}
\usepackage{fcprefix}
\usepackage{amsmath}
\usepackage{amssymb}
\usepackage{array}
\usepackage{bbold}
\usepackage{bm}
\usepackage{booktabs}
\usepackage{float}
\usepackage{graphicx}
\usepackage[utf8]{inputenc}
\usepackage{lipsum}
\usepackage{multirow}
\usepackage{dsfont}
\usepackage{nicefrac}
\usepackage{slashed}
\usepackage[normalem]{ulem}
\usepackage[dvipsnames]{xcolor}

\newcommand\TT{\rule{0pt}{2.0ex}}       % Top strut
\newcommand\BB{\rule[-1.ex]{0pt}{0pt}} % Bottom strut
\usepackage{url,hyperref,lineno,microtype,subcaption}
\usepackage[onehalfspacing]{setspace}
\def\keyFont{\fontsize{8}{11}\helveticabold}
\def\firstAuthorLast{D.\ Sadasivan {et~al.}} %use et al only if is more than 1 author
\def\Authors{D.~Sadasivan\,$^{1,*}$, M.~Mai\,$^{2,4}$, 
M.~D\"oring\,$^{4}$, Ulf-G.~Mei{\ss}ner\,$^{2,3,6}$, 
F.~Amorim\,$^{1}$, J.~Klucik\,$^{1}$, Jun-Xu~Lu$^{6,7}$, 
Li-Sheng~Geng$^{7,8}$}
\graphicspath{{PLOTS/}}
\numberwithin{equation}{section}

\def\Address{
$^{1}$Ave Maria University, Ave Maria, FL 34142, US\\
$^{2}$Helmholtz-Institut f\"ur Strahlen- und Kernphysik (Theorie) and Bethe Center for Theoretical Physics, Universit\"at Bonn, 53115 Bonn, Germany\\
$^{3}$Institute for Advanced Simulation, Insitut f\"ur Kernphysik and J\"ulich Center for Hadron Physics, Forschungszentrum J\"ulich, 52425 J\"ulich, Germany\\
$^{4}$Institute for Nuclear Studies and Department of Physics, The George Washington University, Washington, DC 20052, USA\\
$^{5}$Tbilisi State University, 0186 Tbilisi, Georgia\\
$^{6}$School of Space and Environment, Beihang University, Beijing 102206, China\\
$^{7}$School of Physics, Beihang University, Beijing 102206, China\\
$^{8}$Peng Huanwu Collaborative Center for Research and Education, Beihang University, Beijing 100191, China}
\def\corrAuthor{D.\ Sadasivan}
\def\corrEmail{daniel.sadasivan@avemaria.edu}
\begin{document}
\onecolumn
\firstpage{1}

%%%%%%%%%%%%%%%%%%%%%%%%%%%%%%%
%%%%%%%%%%%%%%%%%%%%%%%%%%%%%%%
\title[New insights into meson-baryon scattering]{
New insights into the pole parameters of the  {\boldmath$\Lambda(1380)$}, the
  {\boldmath$\Lambda(1405)$} and the {\boldmath$\Sigma(1385)$}} 
%%%%%%%%%%%%%%%%%%%%%%%%%%%%%%%
\author[\firstAuthorLast ]{\Authors} %This field will be automatically populated
\address{} %This field will be automatically populated
\correspondance{} %This field will be automatically populated

\extraAuth{}% If there are more than 1 corresponding author, comment this line and uncomment the next one.
%\extraAuth{corresponding Author2 \\ Laboratory X2, Institute X2, Department X2, Organization X2, Street X2, City X2 , State XX2 (only USA, Canada and Australia), Zip Code2, X2 Country X2, email2@uni2.edu}
%%%%%%%%%%%%%%%%%%%%%%%%%%%%%%%
\maketitle

%%%%%%%%%%%%%%%%%%%%%%%%%%%%%%%
%%%%%%%%%%%%%%%%%%%%%%%%%%%%%%%
\begin{abstract}
A coupled-channel S- and P-wave next-to-leading order chiral-unitary approach for strangeness $S=-1$ meson-baryon
scattering is extended to include the new data from the KLOE and AMADEUS experiments as well as the $\Lambda\pi$
mass distribution of the $\Sigma(1385)$. The positions of the poles on the second Riemann sheet
corresponding to the $\Sigma(1385)$ pole and the $\Lambda(1380)$ and $\Lambda(1405)$ poles as well as
the couplings of these states to various channels are calculated. We find that the resonance positions and branching ratios are on average determined with about 20\% higher precision when including the KLOE and AMADEUS data. Additionally, for the first time, the correlations between the parameters of the poles are investigated and shown to be relevant. We also find that the $\Sigma(1385)$ has negligible influence on the properties of the $\Lambda$ states given the available data. Still, we identify isospin-1 cusp structures in the present solution in light of new measurements of $\pi^\pm\Lambda$ line-shapes by the Belle collaboration.
\tiny
%%%%%%%%%%%%%%%%%%%%%%%%%%%%%%%
\keyFont{ \section{Keywords:} Chiral symmetry, coupled channels, strangeness, pole parameters, meson-baryon
scattering, kaonic hydrogen} %All article types: you may provide up to 8 keywords; at least 5 are mandatory.
\end{abstract}
%%%%%%%%%%%%%%%%%%%%%%%%%%%%%%%
%%%%%%%%%%%%%%%%%%%%%%%%%%%%%%%
\section{Introduction}
%%%%%%%%%%%%%%%%%%%%%%%%%%%%%%%

The resonances $\Lambda(1380)1/2^-$, $\Lambda(1405)1/2^-$ and $\Sigma(1385)3/2^+$ dominate low-energy
strangeness $S=-1$ meson-baryon scattering. This region is studied through a variety of methods: chiral
unitary coupled-channel approaches~\cite{Mai:2014xna,Mai:2012dt,Cieply:2016jby,Miyahara:2018onh,Liu:2016wxq,Ramos:2016odk,Kamiya:2016jqc,Feijoo:2015yja,Molina:2015uqp,Guo:2012vv, Doring:2011ip,Ikeda:2011pi,Cieply:2011nq,Cieply:2015pwa,Jido:2002zk,Doring:2010rd,Oller:2006jw,Garcia-Recio:2002yxy,Oset:2001cn,Lutz:2001yb,Oller:2000fj,Oset:1997it,Kaiser:1995cy,Kaiser:1995eg}, amplitude analyses~\cite{Matveev:2019igl, Sarantsev:2019xxm, Anisovich:2020lec,Fernandez-Ramirez:2015tfa, Fernandez-Ramirez:2015fbq, Kamano:2015hxa, Kamano:2014zba, Zhang:2013sva, Zhang:2013cua,Wang:2015qta,Xie:2018gbi}, lattice QCD~\cite{Hall:2016kou, Hall:2014uca, Engel:2013ig, Edwards:2012fx, Engel:2010my}, and quark models~\cite{Xie:2014zga, Helminen:2000jb, Jaffe:2003sg}, see, e.g., the reviews in Refs.~\cite{Hyodo:2011ur,Guo:2017jvc,Tanabashi:2018oca,Meissner:2020khl} and the recent review dedicated to the $\Lambda(1405)$~\cite{Mai:2020ltx}. 
We highlight the recent effort to simultaneously analyze the three strangeness $S=\pm1 ,0$ sectors with a next-to-next-to-leading order (NNLO) amplitude in unitarized chiral perturbation theory~\cite{Lu:2022hwm}.

Knowledge of the two $\Lambda$ states provides
insight into the generation of the $K^-pp$ bound state~\cite{Ajimura:2018iyx,Sada:2016nkb} 
as demonstrated in
Ref.~\cite{Sekihara:2016vyd} and into neutron stars, whose equation of state is sensitive to the
propagation of antikaons via the behavior of antikaon condensate~\cite{Kaplan:1986yq, Pal:2000pb}.
For recent reviews discussing these aspects, see e.g. Refs.~\cite{Tolos:2020aln,Hyodo:2022xhp}.

When the scattering amplitude is analytically continued to the second Riemann sheet, the poles of the
$\Lambda(1380)$, the $\Lambda(1405)$ and the $\Sigma(1385)$ can be observed. Note that we have already
made explicit the remarkable two-pole structure in the region of the  $\Lambda(1405)$, which was first
observed in the context of chiral-unitary approaches in Ref.~\cite{Oller:2000fj} and is now reflected in the
listings in the Review of Particle Physics~\cite{ParticleDataGroup:2022pth} (though not yet in the summary
tables). For a general review on such two-pole structures in QCD, see~\cite{Meissner:2020khl}. Coming back to the poles under consideration, the amplitude can be uniquely described by the complex pole positions and  residues, which are determined by fitting models to data. The uncertainties on the pole predictions and residues can be constrained by recently measured data from AMADEUS~\cite{Piscicchia:2018rez} and KLOE~\cite{Piscicchia:2022wmd}.
Studying the impact of these new data on the chiral unitary amplitudes and resonance poles is the main motivation of this paper.
In addition, we investigate the influence of the $\Sigma(1385)$. This resonance is sub-threshold with respect to the $\bar K N$ channel and there is a centrifugal barrier due to its P-wave nature. Yet, it is not far from the $\bar KN$ threshold and could have an influence on low-energy $\bar K N$ data through its finite width. To estimate the influence, we include the line-shape data from Ref.~\cite{Baubillier:1984pj} in the analysis that warrants a physical mass and width of the $\Sigma(1385)$.

Furthermore, the predictions of the pole positions and residues of the $\Lambda(1380)$, the  $\Lambda(1405)$ and the $\Sigma(1385)$ are correlated. Quoting correlations is as relevant as quoting error bars to confine the uncertainty region more meaningfully. For the first time, we calculate the pole correlations for a meson-baryon system.

This manuscript is organized as follows: In Sect.~\ref{sec:formalism} we briefly discuss the underlying
coupled-channel approach that is used to analyze the data. The fit to the available data from
antikaon-proton scattering,  kaonic hydrogen and the so-called threshold ratios are displayed in Sect.~\ref{sec:fits}. 
The investigation of the correlations between the various pole parameters is presented in Sect.~\ref{sec:pole-correlations}, followed by the study of the
impact of the new data from KLOE and AMADEUS on the pole positions of the
$\Lambda(1380)$, the $\Lambda(1405)$ and the $\Sigma(1385)$.
In Sec.~\ref{sec:belle} we discuss the current solution in light of the new $\pi^\pm\Lambda$ line-shape measurements by the Belle collaboration~\cite{Belle:2022kvv}.
We end with a summary and discussion in Sect.~\ref{sec:summ}.
Some further results are displayed in the appendix.

%%%%%%%%%%%%%%%%%%%%%%%%%%%%%%%
\section{Formalism}
\label{sec:formalism}
%%%%%%%%%%%%%%%%%%%%%%%%%%%%%%%
In this work we use an approach derived in a series of
works~\cite{Bruns:2010sv,Mai:2013cka,Mai:2014xna} which has the correct low-energy behavior by including all contact interactions from the leading (LO) and next-to-leading (NLO) chiral Lagrangian, while it also fulfills two-body unitarity. The latter issue is crucial for two reasons: first, it allows one to formally access the resonance parameters from poles on the second Riemann sheet; secondly, the re-summation of the interaction kernel allows to extend the applicability region of the approach, which indeed spans several hundred MeV in the present case. The downside is that the re-summation procedure is not unique and, thus, some model-dependence is introduced, with the corresponding parameters being determined from experimental data. Still, in a given scheme the procedure is systematically improvable by including kernels of higher order as being performed recently, see Ref.~\cite{Lu:2022hwm}.
%The non-perturbative amplitude for strangeness $S=-1$ meson-baryon scattering from this approach, which is valid over a large energy range including the resonance region, requires some resummation technique. Thus, some model-dependence is introduced, with the corresponding parameters being fitted to experimental data. 
Finally, we note that since the underlying degrees of freedom are the members of the ground state meson and the ground state baryon octet, the $\Lambda(1380)$, $\Lambda(1405)$ % and $\Sigma(1385)$ 
resonances are dynamically generated without being explicitly introduced, so that their existence and properties can be considered as genuine predictions.

In the following we recap the main steps in accessing observables and relating them to the resonance parameters. The $T$-matrix is defined in terms of the $S$-matrix as $S=\mathds{1}-iT$. The corresponding meson-baryon scattering amplitude for the process $M(q_1)B(p-q_1)\to M(q_2)B(p-q_2)$ is then a spinor function $T(\slashed{q}_2, \slashed{q}_1; p)$, where total four-momentum $p$ conservation is already assumed. This quantity can now be conveniently derived from the three-flavour CHPT Lagrangian~\cite{Krause:1990xc,Frink:2004ic}
%%%%%%%%%%%%%
\begin{align}
&T_{\rm LO}(\slashed{q}_2, \slashed{q}_1; p)=A_{WT}(\slashed{q_1}+\slashed{q_2})\,,\nonumber\\
&T_{\rm NLO}(\slashed{q}_2, \slashed{q}_1; p)=A_{1-4}(q_1\cdot q_2)+A_{5-7}[\slashed{q_1},\slashed{q_2}]
+A_{0DF} +A_{8-11}\Big(\slashed{q_2}(q_1\cdot p)+\slashed{q_1}(q_2\cdot p)\Big)\,
\label{eq:potential}
\end{align}
%%%%%%%%%%%%% 
for a Minkowski four-momentum product $(x\cdot y)$ and commutator $[a,b]=ab-ba$. Here the momentum/spinor structures are conveniently separated  from the channel-space matrices $A$ as encoded in the chiral Lagrangian. Specifically for  strangeness $S=-1$, we have $10\times10$ real-valued matrices with respect to the channels $\mathcal{S}:=\{K^-p$, $\bar K^0 n$, $\pi^0\Lambda$, $\pi^0\Sigma^0$, $\pi^+\Sigma^-$, $\pi^-\Sigma^+$, $\eta\Lambda$, $\eta \Sigma^0$, $K^+\Xi^-$, $K^0\Xi^0\}$, see the Appendix of Ref.~\cite{Mai:2014xna} for explicit formulae.

So far, the usage of CHPT has allowed us to put constraints on possible momentum and flavour structures of the scattering amplitude~\eqref{eq:potential}. Including this into the so-called chiral unitary approach is done by utilizing the Bethe-Salpeter equation in $d$ space-time dimensions in Minkowski space,
%%%%%%%%%%%%%
\begin{eqnarray}
\label{eqn:BSE}
T^{ij}(\slashed{q}_2,\slashed{q}_1; p)&=
&V^{ij}(\slashed{q}_2, \slashed{q}_1;p)
+
i\int\frac{d^d \ell}{(2\pi)^d}
\frac{V^{ik}(\slashed{q}_2, \slashed{\ell}; p)}{\ell^2-M_k^2+i\epsilon}
\frac{1}{\slashed{p}-\slashed{\ell}-m_k+i\epsilon}
T^{kj}(\slashed{\ell}, \slashed{q}_1; p)\,,\nonumber\\
&&\text{for~~~}
V(\slashed{q}_2, \slashed{q}_1;p):=
T_{\rm LO}(\slashed{q}_2, \slashed{q}_1; p)+
T_{\rm NLO}(\slashed{q}_2, \slashed{q}_1; p)\,,
\end{eqnarray}
%%%%%%%%%%%%%
where $i,j,k\in\mathcal{S}$ and $m/M$ are
the mass of the baryon/meson in each channel, respectively. The interaction kernel $V$ of the above integral equation~\eqref{eqn:BSE} is restricted to the contact terms only, i.e., it neglects the presence of the baryon exchange diagrams, the so-called Born-terms. In general, such terms lead to more complex analytical structures, e.g., left-hand cuts in various coupled channels, see e.g. the discussion in~\cite{Meissner:1999vr}
While the solution of Eq.~\eqref{eqn:BSE} is not known in such a case, it can be solved analytically (see Refs.~\cite{Bruns:2010sv, Mai:2012dt}) when only contact terms are taken into account. For more details on this issue and comparison to other approaches, see the review~\cite{Mai:2020ltx}. The UV-divergence inherent to Eq.~\eqref{eqn:BSE} is tamed by dimensional regularization in the $\overline{\rm MS}$ scheme, which introduces a regularization scale. While the natural size of this scale is discussed in Ref.~\cite{Oller:2000fj}, we note that in the present model it accounts for the Feynman topologies not included  by the Bethe-Salpeter equation. The scales are, therefore, regarded as  free parameters channel-by-channel and referred henceforth to as $\{a_i|i=1,..,6\}$, neglecting isospin breaking. These parameters accompany low-energy constants (LECs) $\{b_0,b_D,b_F,b_1,...,b_{11}\}$ parametrizing matrices $A_{1-4},A_{0DF},A_{5-7},A_{8-11}$, respectively, as the free parameters of this model. Note that the leading-order Weinberg-Tomozawa amplitude
$A_{WT}$ only depends on the pseudoscalar meson decay constant, which we fix together with all relevant hadron masses to their physical values.

Having defined the scattering amplitude, we obtain partial waves in the standard way~\cite{Hohler:1984ux}. Specifically, the partial-wave amplitudes for a transition $i\to j$ reads
%%%%%%%%%%%%%
\begin{align}
\label{eq:famplitudes}
f_{L\pm}^{ij}=
\frac{\sqrt{E_i+m_i}\sqrt{E_j+m_j}}{16\pi W }&\left(A_L^{ij}+\left(W-\frac{m_i+m_j}{2}\right)B_L^{ij}\right)\\
&~~~~-\frac{\sqrt{E_i-m_i}\sqrt{E_j-m_j}}{16\pi W }\left(A_{L\pm1}^{ij}-\left(W+\frac{m_i+m_j}{2}\right)B_{L\pm1}^{ij}\right)\,,\nonumber
\end{align}
%%%%%%%%%%%%%
where $W=p^0$ is the total energy in the center-of-mass system (CMS), $L_{\pm}:=L\pm1/2$ is the total angular momentum, the relative angular momentum is $L$, the modulus of the three-momentum in the CMS is $q_{cms,i}$ and  $E_i:=\sqrt{m_i^2+q_{cms,i}^2}$. Finally, the quantities $A_L^{ij}$ and $B_{L}^{ij}$ are the partial-wave projected
invariant amplitudes, obtained from the scattering amplitude~\eqref{eqn:BSE} as
$T^{ij}_{\rm on-shell}=A^{ij}+(\slashed{q}+\slashed{q'})B^{ij}$. Note that we neglect the Coulomb interaction
in the scattering processes involving charged particles.

The  definition~\eqref{eq:famplitudes} shows the relation between partial waves and momentum structures of the scattering amplitude~\eqref{eqn:BSE}. This leads to an interesting observation discussed in Refs.~\cite{Sadasivan:2018jig,Mai:2012wy} that because the momentum structures are truncated as shown in Eq.~\eqref{eq:famplitudes} both $f_{0+}$ and $f_{1-}$ partial-waves are indeed complete in the sense that all partial-wave amplitudes $A_L^{ij}$ and $B_{L}^{ij}$ required for their calculation are taken into account. Besides the pioneering work of the Munich group~\cite{CaroRamon:1999jf} this approach
is the only existing unitary coupled-channel model which contains explicit S- and P-wave interactions,
derived from the low-energy behavior of QCD Green's functions. In contrast, $f_{1+}$ can only be partially reconstructed as it lacks $A_2^{ij}$ and $B_{2}^{ij}$ terms. This presents a challenge for predicting the
pole position of the $\Sigma(1385)3/2^+$, but is overcome in Ref.~\cite{Sadasivan:2018jig} using the two-potential
formalism~\cite{Ronchen:2012eg}. It allows one to include an explicit resonance to an existing unitary approach without spoiling unitarity. There, we incorporate the $\Sigma(1385)$, modifying the isovector $f_{1+}$ amplitude using the two-potential formalism~\cite{Ronchen:2012eg} extrapolated into the sub-threshold region as
%%%%%%%%%%%%%
\begin{align}
f_{1+}\mapsto f_{1+}+f_{1+}^P\,,
&\text{~~~for~~~}f_{1+}^{P,ij}=\Gamma^i\Gamma^{j}\,\left(W-m_\Sigma^0-\sum_{k=1}^3\gamma_k I_{MB,k} \Gamma^k\right)^{-1}\nonumber\\
&\text{~~~with~~}\Gamma^i= \gamma^i+  \sum_{k=1}^3\gamma_k I_{MB,k} f_{1+}^{ki}\,,
\label{eq:2pot}
\end{align}
%%%%%%%%%%%%%
for $i,j,k\in\{\pi\Lambda,\pi\Sigma,\bar{K}N\}|_{I=1}$. Here, $I_{MB,k}$ is the meson-baryon loop function~\cite{Mai:2012wy}, whereas ${\gamma_i=q_{cms,i}\lambda\,\lambda_i}$ is the ``bare vertex'' with one free fit parameter $\lambda$ and the relative decay strengths $\lambda_i$ to channels $\pi\Lambda$, $\pi\Sigma$, and $\bar{K}N$ fixed by the Lagrangian of Ref.~\cite{Butler:1992pn}, see also Ref.~\cite{Doring:2005bx}. This is due to the fact that the available data on the $\Sigma(1385)$ cannot individually resolve these channels~\cite{Kim1966columbia}. The bare mass $m_\Sigma^0$ and
bare coupling $\lambda$ are new fit parameters. Additionally, we include a factor $f_\Sigma$ to scale the final-state $\pi\Lambda \to \pi\Lambda$ interaction to the process in the experiment~\cite{Baubillier:1984pj}. 

%%%%%%%%%%%%%%%%%%%%%%%%%%%%%%%
\section{Results}
\label{sec:results}
%%%%%%%%%%%%%%%%%%%%%%%%%%%%%%%

%%%%%%%%%%%%%%%%%%%
%%%%%%%%%%%%%%%%%%%
\begin{table*}[t]
\scriptsize
% \hspace{0cm}
\addtolength{\tabcolsep}{-1mm}
\renewcommand{\arraystretch}{1.07}
\centering
\begin{tabular}{|l|r|r|r|r|r|p{3mm}|}
\cline{1-6}
%%%%%%%%%%%
Observable
&\# data & $\mathfrak{F}_1$ (all new data) &  $\mathfrak{F}_2$ (new amp) &  $\mathfrak{F}_3$ (new cs) &  $\mathfrak{F}_4$ (no new data)&
\multicolumn{1}{c}{} \TT\\
\hline
SIDDHARTA
& 2 & 2.09 & 2.24 &  1.57 & 1.06&\multirow{ 11}{*}{\rotatebox{90}{\tiny Old data}~~}\\
$\gamma,R_c,R_n$ & 3 & 2.15 & 0.38 &  1.66 &  0.10& \\
$\sigma_{K^-p\to K^-p}$& 32 & 56.60 & 63.71 &  60.22 & 69.15 &\\
$\sigma_{K^-p\to\bar K^0n}$& 37 & 66.57 & 63.52 &  66.87 & 70.19&  \\
$\sigma_{K^-p\to\pi^+ \Sigma^-}$& 39 & 50.32 & 42.16 &  46.39 & 35.99& \\
$\sigma_{K^-p\to\pi^- \Sigma^+}$& 41 & 82.63 & 65.52 &  72.93 & 55.95 &\\
$\sigma_{K^-p\to\pi^0 \Lambda}$& 3 & 0.80 & 0.30 &  1.21 & 0.24&\\
$\sigma_{K^-p\to\pi^0 \Sigma^0}$& 3 & 0.42 & 0.26 &  0.50 & 1.01&\\
$d\sigma/d\Omega(K^-p\to K^-p)$& 153 & 311.86 & 326.24 &  327.93 & 351.40& \\
$d\sigma/d\Omega(K^-p\to\bar K^0 n)$&60 & 73.63 & 71.06 &  72.93 & 69.97&\\
% \cline{2-7}
$\Sigma (1385)$ line-shape~\cite{Baubillier:1984pj} & 38 & 28.42 & 26.07 & 28.377 & 24.89 & \\
\hline
$\sigma_{K^-p\to\pi^0 \Lambda}$~\cite{Piscicchia:2022wmd}
& 1 & 0.63 & (19.18) &  0.89 & (15.20)&
\multirow{ 3}{*}{\rotatebox{90}{\tiny New data}~~}\TT\\
$\sigma_{K^-p\to\pi^0 \Sigma^0}$~\cite{Piscicchia:2022wmd}
& 1 & 1.78 & (3.74) &  1.76 & (5.08)&\\
$|f_{0+}^{\pi^-\Lambda \to K^-n}|$~\cite{Piscicchia:2018rez}
& 1 & 0.04 & 0.02 &  (2.88) & (5.63)&\BB\\
\hline
\multicolumn{2}{c|}{}&&&&&\multicolumn{1}{c}{}\\[-0.3cm]
%%%%%%%%%%%
\multicolumn{2}{r|}{$\sum_{a}\chi^2_{a}$}& 677.98 & 687.46 &  685.00 & 705.92&\multicolumn{1}{c}{}\\
\multicolumn{2}{r|}{$\chi^2_{\rm dof}$}& 1.19 & 1.06 &  1.26 & 1.13&\multicolumn{1}{c}{} \\
\cline{3-6}
\multicolumn{7}{c}{}\\[-0.3cm]
\cline{3-6}
\multicolumn{2}{r|}{$W^*_{\Lambda(1405)}\,[{\rm GeV}]$}&$(1.430-0.023i)$ & $(1.431 -0.029i)$ &  $(1.431-0.018i)$ & $(1.427-0.017i)$&\multicolumn{1}{c}{}\TT\\
\multicolumn{2}{r|}{$W^*_{\Lambda(1380)}\,[{\rm GeV}]$}&$(1.355-0.036i)$ & $(1.300-0.019i)$ &  $(1.346-0.029i)$ & $(1.347-0.027i)$&\multicolumn{1}{c}{}\\
\multicolumn{2}{r|}{$W^*_{\Sigma(1385)}\,[{\rm GeV}]$} & $(1.385-0.019i)$ & $(1.385-0.019i)$ &  $(1.385-0.020i)$ & $(1.383-0.019i)$
&\multicolumn{1}{c}{}\\
%%%%%%%%%%%
\cline{3-6}
\end{tabular}
\caption{
\label{tab:chi2-all}
Individual and total $\chi^2$ for the fit strategy $\mathfrak{F}_1,\dots, \mathfrak{F}_4$. The individual contributions to the $\chi^2$ are the $\chi^2_a$ which contributes to the $\chi^2_\text{dof}$ as in Eq.~\eqref{eq:chi2un}. Predicted observables (not included in $\chi^2$) are put in parentheses. Bottom part of the table collects the predicted pole positions $W^*\in\mathds{C}$. Uncertainties on pole positions are shown separately.
}
\end{table*}
%%%%%%%%%%%%%%%%%%%
%%%%%%%%%%%%%%%%%%%
%%%%%%%%%%%%%%%%%%%%%%%%%%%%%%%
\subsection{Fits}
\label{sec:fits}
%%%%%%%%%%%%%%%%%%%%%%%%%%%%%%%

The fits performed here represent a considerable step forward for two reasons. First, because the model confronts  highly anticipated, recently measured data from the AMADEUS~\cite{Piscicchia:2018rez} and KLOE~\cite{Piscicchia:2022wmd} collaborations. These data consists of $|f_{0+}^{\pi^-\Lambda \to K^-n}|$ at $W\approx1400{\rm MeV}$ and $\{\sigma_{K^-p\to\pi^0\Lambda}$, $\sigma_{K^-p\to\pi^0\Sigma^0}\}$ at $W\approx1438{\rm MeV}$, respectively. Second, we study the impact of the older data from Ref.~\cite{Baubillier:1984pj} on the invariant mass distribution for the $(\Lambda\pi^+)$ final state in the $K^-p\to(\Lambda\pi^+)\pi^-$ reaction. To our knowledge this data has not been considered before in this context. In addition, we also include the following, previously considered data:
%%%%%%%%%%%%%
\begin{itemize}

\item The six channels with available total cross section data for $I(J^P)=0(\frac{1}{2}^-)$, $S=-1$ meson-baryon interaction with thresholds close enough to sizeably contribute to the $\bar{K}N$ amplitude around its threshold: $K^-p\to K^-p$, $K^-p\to\bar{K}^0n$,
$K^-p\to\pi^0\Lambda$,
$K^-p\to\pi^0\Sigma^0$, $K^-p\to\pi^+\Sigma^-$, $K^-p\to\pi^-\Sigma^+$~\cite{Ciborowski:1982et,Humphrey:1962zz,Sakitt:1965kh,Watson:1963zz}.

\item The differential cross section data for the $K^-p\rightarrow K^-p$ and $K^-p\rightarrow \bar K^0n$ channels~\cite{Mast:1975pv} with energies where the CHPT kernel is a good approximation.

\item The measurements of the energy shift and width of kaonic hydrogen performed by the SIDDHARTA collaboration, see Ref.~\cite{Bazzi:2011zj}. These are related in Ref.~\cite{Meissner:2004jr} to the complex $\bar{K}N$ scattering lengths at the threshold including isospin breaking.

\item The decay ratios $\gamma=(K^-p \rightarrow \Sigma^-\pi^+)/(K^-p \rightarrow \Sigma^+\pi^-)$, $R_n=(K^-p \rightarrow \Lambda \pi^0)/(K^-p \rightarrow \Lambda\pi^0,\Sigma^0\pi^0)$, $R_c=(K^-p \rightarrow \text{charged  particles})/(K^-p \rightarrow \text{all final states})$ from Refs.~\cite{Tovee:1971ga,Nowak:1978au}. All ratios are taken at the $K^-p$ threshold.

\end{itemize}
%%%%%%%%%%%%%
The summary of all considered data can be found in Tab.~\ref{tab:chi2-all}. Note that the old data are discussed in detail in the dedicated review~\cite{Mai:2020ltx} including links to an open GitHub repository containing these data in sorted, digital form.

In order to isolate the impact of the recently measured data in comparison to that of the established data set, we consider four different data fit scenarios. Scenario $\mathfrak{F}_1$ includes all data discussed above, i.e., old and new ones from Refs.~\cite{Piscicchia:2018rez,Piscicchia:2022wmd}. Scenario $\mathfrak{F}_2$ includes the same data except the KLOE data~\cite{Piscicchia:2022wmd}. Scenario $\mathfrak{F}_3$ includes all data except the AMADEUS data~\cite{Piscicchia:2018rez}. Case $\mathfrak{F}_4$ includes all of the older data but neither of the recent measurements~\cite{Piscicchia:2018rez,Piscicchia:2022wmd}. For each of these cases, the weighted $\chi^2$ according to
%%%%%%%%%%%%%
\begin{align}
\chi^2_{\rm dof}=\frac{\sum_a N_a}{A((\sum_a N_a)-n)}
\sum_{a=1}^{A}
\frac{\chi^2_a}{N_a}
\text{~~~with~~~}
\chi^2_a=
\sum_{i=1}^{N_a}
\left(\frac{f^a_i(\vec\aleph)-\hat f^a_i}{\Delta \hat f^a_i}\right)^2
%\chi^2_{\rm penalty}&=3(g-4.98)^2+3(m_\Sigma^0-3.23)^2\,
\label{eq:chi2un}
\end{align}
%%%%%%%%%%%%%%%%%%%%%%
is minimized with respect to $n=23$ free parameters collected in the vector $\vec\aleph =(a_1,..a_6,b_0,b_
D,b_F,b_1,..,b_{11},m_\Sigma^0,\lambda,f_\Sigma)$. The number of data for an observable $a\in\{1,..,A\}$ is denoted by $N_a$, and $\hat f^a_i$ are the data with uncertainties $\Delta \hat f^a_i$. The present choice of $\chi^2_{\rm dof}$ takes account of the very unequal distribution of number of data points in different observables, giving more weight to observables with fewer data.

Our fitting procedure involves finding the parameters for case $\mathfrak{F}_1$ by minimizing $\chi^2_{\rm dof}$ starting with randomly generated free parameters. We found one set of parameters had a $\chi^2_{\rm dof}$ an order of magnitude smaller than all other $\chi^2_{\rm dof}$, comprising our fit result $\mathfrak{F}_1$. Subsequently, we use these parameters as starting parameters for the minimization of $\chi^2_{\rm dof}$ for each other scenarios. The result of this procedure for all fit scenarios is summarized in Tab.~\ref{tab:chi2-all}, while the best fit parameters are relegated to the appendix, see Tab.~\ref{tab:parameters}.

In summary, we observe that the new data~\cite{Piscicchia:2018rez, Piscicchia:2022wmd} do indeed provide a non-negligible constraint on the coupled-channel formalism, e.g., individual contributions  $\chi^2_{a}/N_a$ of these data are substantially larger than those of the older data, see $\mathfrak{F}_{2-4}$. There is, however, enough elasticity in the current chiral unitary approach, providing an adequate description of all data, see $\mathfrak{F}_1$. A more detailed discussion of the impact of the new data on the various fits and their results is provided below.

%%%%%%%%%%%%%%%%%%%%%%%%%%%%%%%
\subsection{Amplitudes and poles}
\label{sec:pole-correlations}
\begin{figure}[htbp]
    \includegraphics[width=0.49\linewidth]{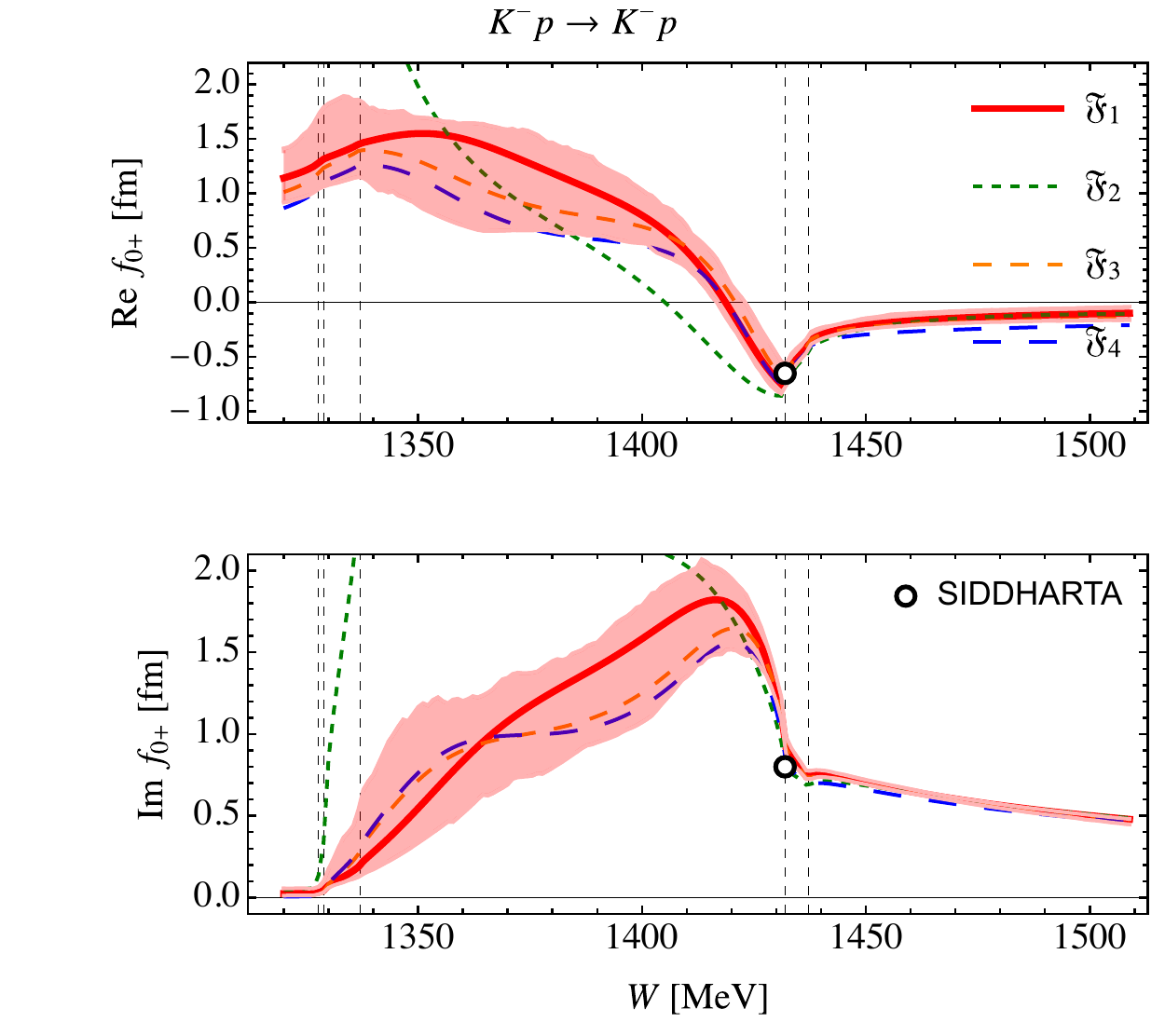}
    \includegraphics[width=0.49\linewidth]{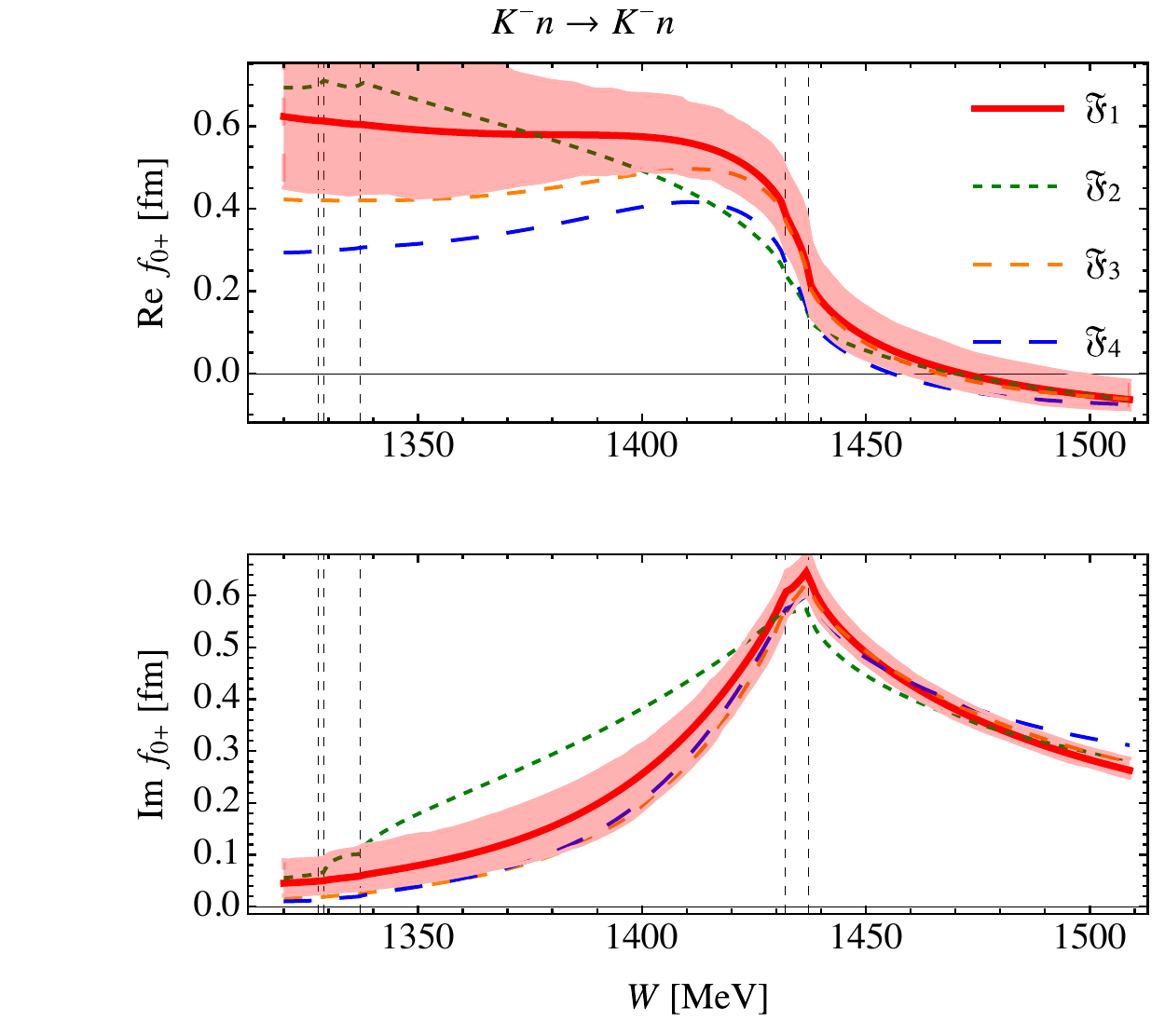}
    \caption{Scattering amplitudes for physical channels. Here, the $K^-n\to K^-n$ amplitude is determined assuming isospin invariance. Vertical dashed lines represent the positions of the relevant two-body thresholds.}
    \label{fig:KbarNphys}
\end{figure}

%%%%%%%%%%%%%%%%%%%%%%%%%%%%%%%
The scattering amplitudes for the $K^-p\to K^-p$ transition in  S-wave is shown in the left column of Fig.~\ref{fig:KbarNphys} for the four fit scenarios. For fit $\mathfrak{F}_1$ that contains all new data, we determine the statistical $1\sigma$ uncertainty region through re-sampling. In that, we first perform a fit to the original data. Then, the data is varied randomly with respect to provided statistical uncertainties and a new fit starting with the original one is performed. This procedure is then repeated sufficiently many times, and is done for each fit scenario. However, we refrain from showing the resampling for the other fits to keep the figures simple. As the figure shows, the amplitude is not very sensitive to (ex-)inclusion of the new data from Refs.~\cite{Piscicchia:2018rez,Piscicchia:2022wmd} within statistical uncertainties except for $\mathfrak{F}_2$ that is very different.
%%%%%%%%%%%%%%%%%%%%%%
%%%%%%%%%%%%%%%%%%%%%%
%%%%%%%%%%%%%%%%%%%%%%
%%%%%%%%%%%%%%%%%%%%%%
The less known $K^-n\to K^-n$ amplitude, shown in the right column, shows comparable variations, especially below the $\bar KN$ thresholds.  This is also the region where the sub-threshold AMADEUS data~\cite{Piscicchia:2018rez} is measured.  All the partial waves for $\bar KN$ scattering  in both isospin channels are collected in Fig~\ref{fig:KbarNIsospin} in the Appendix. For the P-waves, we observe a similar pattern as for the S-waves: The fits $\mathfrak{F}_2$, $\mathfrak{F}_3$, and $\mathfrak{F}_4$ stay within the uncertainty band of $\mathfrak{F}_1$ up to slightly larger deviations in some occasions. In Fig.~\ref{fig:KbarNIsospin} to the upper left we also observe the superposition of the $\Lambda(1380)$ and $\Lambda(1405)$ poles. Still, the $\Sigma(1385)$ couples very weakly to the $\bar{K}N$ channels and its effect is unresolved in the $\bar KN$ amplitudes but can be observed distinctly in other channels, such as $\pi^0 \Lambda \rightarrow \pi^0 \Lambda$. We also observe a considerable influence of the $\Sigma(1385)$ in the $\pi\Sigma$ channel. As we fit $\pi\Sigma$ data with mixed isospin, changes in $I=1$ amplitude modify the $I=0$ amplitude (to get the same data description). This explains that despite having different isospin, the $\Sigma(1385)$ has some limited influence on the $\Lambda(1380)$ parameters as discussed below.

\begin{figure}[htbp]
\includegraphics[height=5.4cm]{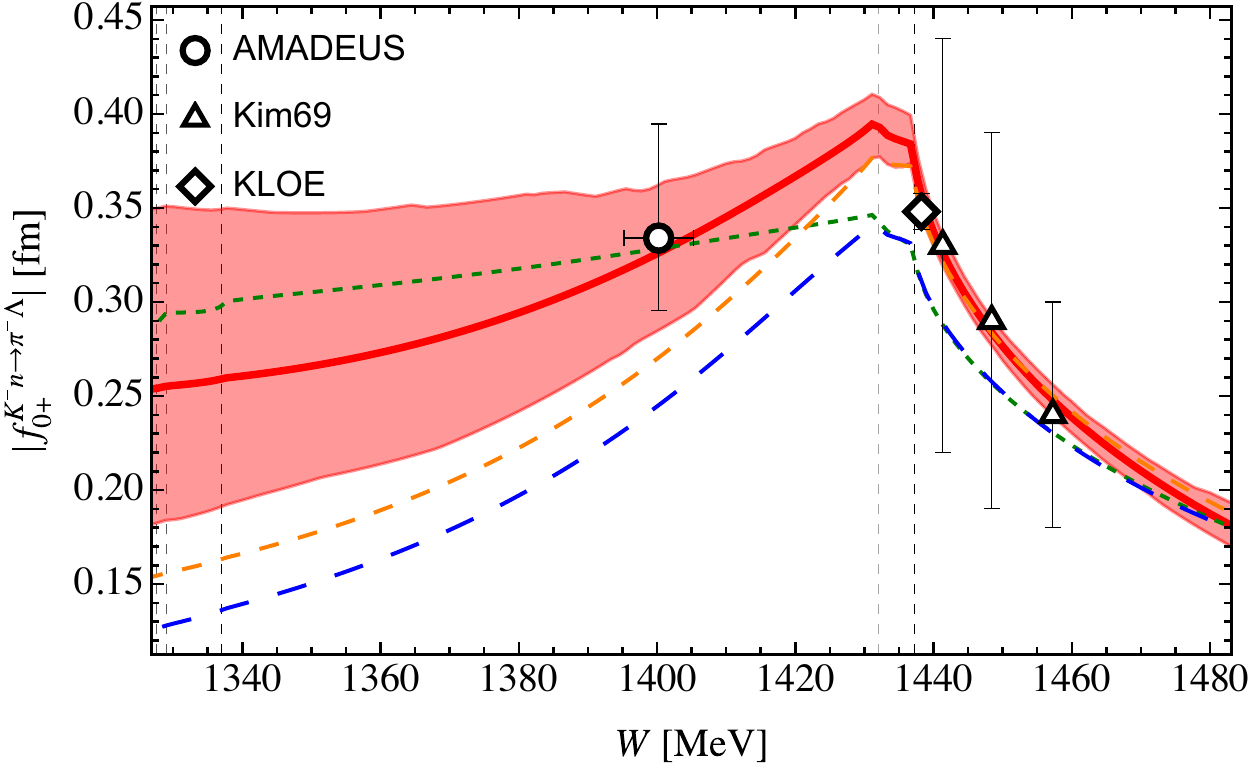}
\includegraphics[height=5.4cm]{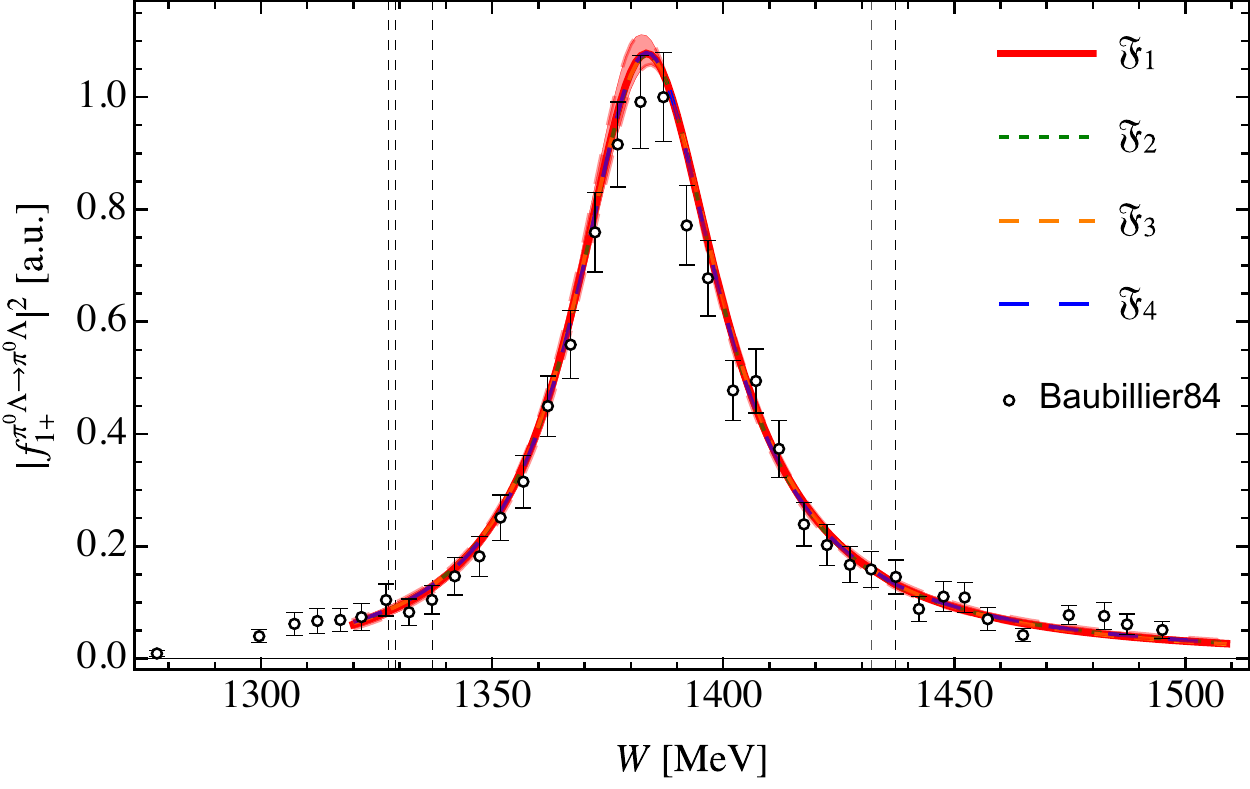}
    \caption{Comparison of the best fit results to the 
    new data. Different line-shapes correspond to fit strategies $\mathfrak{F}_{1-4}$ with $1\sigma$ band plotted only for the all-data fit $\mathfrak{F}_1$. Experimental results are represented by the black empty symbols referring to Baubillier84~\cite{Baubillier:1984pj}, AMADEUS~\cite{Piscicchia:2018rez}, KLOE~\cite{Piscicchia:2022wmd} and Kim69~\cite{Kim1966columbia}. The latter two measure cross sections and are included as such into the corresponding fits. In the figure, those are used to estimate partial-waves amplitudes (assuming isospin symmetry and S-wave dominance) to guide  the eye.}
\label{fig:Amplitudes}
\end{figure}

In regard of the amplitudes with $\pi \Lambda$ final states, the result of all four fit scenarios is shown in Fig.~\ref{fig:Amplitudes}. There, we also include data points calculated from the total cross section data~\cite{Kim1966columbia,Piscicchia:2022wmd} assuming S-wave dominance and isospin symmetry. We emphasize that this is only done to guide the eye, all relevant fits include this data as cross sections directly. In the right panel of the same figure we show the results of the line-shape in the $\pi\Lambda\to\pi\Lambda$ channel compared to the data from Ref.~\cite{Baubillier:1984pj}.  We observe no statistically noteworthy impact of the inclusion of the new data~\cite{Piscicchia:2018rez,Piscicchia:2022wmd} on the $\pi\Lambda$ line-shape. However, the $K^-p \to \pi^0 \Lambda$ amplitude does change significantly when including these data. Especially, the datum by the AMADEUS collaboration~\cite{Piscicchia:2018rez} does have a dramatic effect.
%%%%%%%%%%%%%%%%%%%%%%%%
%%%%%%%%%%%%%%%%%%%%%%%%
%%%%%%%%%%%%%%%%%%%%%%%%
%%%%%%%%%%%%%%%%%%%%%%%%

\begin{figure}[htbp]
    \begin{center}
    \hspace{-1cm}\includegraphics[width=0.8\linewidth]{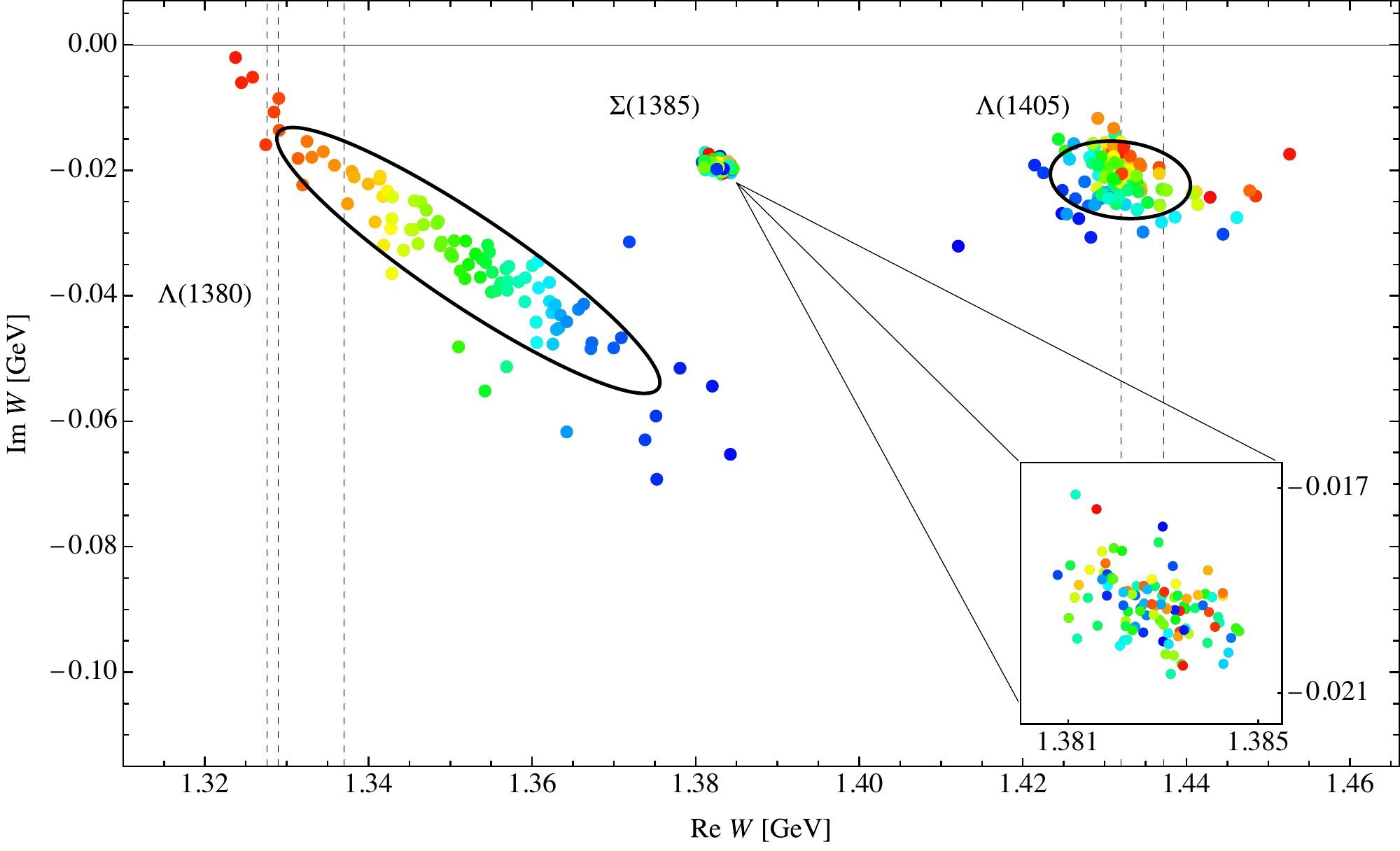}\\
    \includegraphics[width=\linewidth]{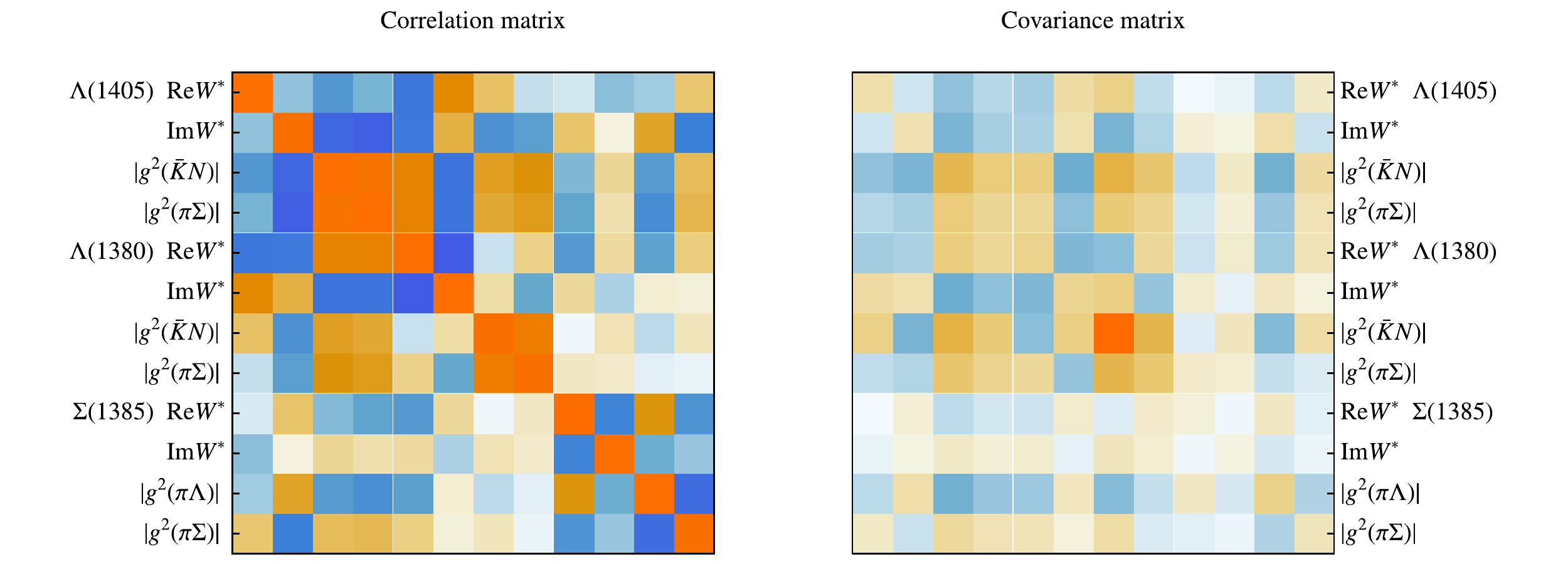}
    \end{center}
    \caption{Summary of the all-data fit ($\mathfrak{F}_1$) results with respect to predicted resonance parameters. {\bf Top}: Confidence regions of the pole positions. Each point corresponds to a pole from a re-sampling solution. Points of the same hue belong to the same sample. The ellipses show $1\sigma$ uncertainty regions of each pole. {\bf Bottom}: Covariance and correlation matrix between pole parameters. These parameters are the real and imaginary positions of each pole as well as the couplings to channels. Redder pixels correspond to more positive values and bluer pixels to more negative values.}
    \label{fig:covcormat}
\end{figure}

Our fits lead to poles on the second Riemann sheet corresponding to the $\Lambda(1380)$, $\Lambda(1405)$ and the $\Sigma(1385)$-resonances. The coupling of these resonances to a meson-baryon channel $i$ is extracted using following expansion $f^{ii}_{L\pm}(W)=g^2_i/(W-W^*)+{\mathcal O}(W^0)$ with $W^*$ being the resonance pole position. The quality of the fits and central results for the pole positions of the four fit scenarios are given in Tab.~\ref{tab:chi2-all} while best fit parameters are relegated to the appendix, see Tab.~\ref{tab:parameters}. These parameters define a scattering amplitude which satisfies an analyticity constraint -- eschewing poles on the first Riemann sheet. In practice, we do not search for poles farther than $150~{\rm MeV}$ from the real axis. The uncertainties of the pole positions are determined in a re-sampling routine. A detailed analysis of the re-sampled points is given in Figs.~\ref{fig:covcormat} and \ref{fig:Bootstrap}.
Our central result -- fit scenario $\mathfrak{F}_1$ corresponding to all-data fit -- yields the following predictions for the pole positions and couplings
%%%%%%%%%%%%%%
%%%%%%%%%%%%%%
\begin{align}
\label{eq:F1-L1405-pole-coupling}
&W^*_{\Lambda(1405)}=1.430(6)\phantom{6}-i\,0.023(4)\phantom{6}\;{\rm GeV}
~~~~&&
g^2={\footnotesize
\begin{pmatrix}
-0.101(98)-i0.193(69)&K^-p\\-0.090(98)-i0.171(60)&\bar{K}^0n\\+0.048(24)+i0.039(29)&\pi^0\Sigma^0\\+0.055(26)+i0.036(31)&\pi^+\Sigma^-\\+0.041(23)+i\,0.040(26)&\pi^-\Sigma^+
\end{pmatrix}
}
\end{align}
%%%%%%%%%%%%%%
%%%%%%%%%%%%%%
\begin{align}
\label{eq:F1-L1380-pole-coupling}
&W^*_{\Lambda(1380)}=1.355(16)-i\,0.038(14)\;{\rm GeV}
~~~~&&
g^2={\footnotesize
\begin{pmatrix}
-0.038(209)+i0.146(135)&K^-p\\
-0.036(147)+i0.144(209)&\bar{K}^0n\\
-0.110(37)\phantom{6}+i0.103(56)\phantom{6}&\pi^0\Sigma^0\\
-0.118(36)\phantom{6}+i0.102(55)\phantom{6}&\pi^+\Sigma^-\\
-0.102(38)\phantom{6}+i0.101(54)\phantom{6}&\pi^-\Sigma^+
\end{pmatrix}
}
\end{align}
%%%%%%%%%%%%%%
%%%%%%%%%%%%%%
\begin{align}
\label{eq:F1-L1385-pole-coupling}
&W^*_{\Sigma(1385)}=1.385(1)\phantom{6}-i\,0.019(1)\phantom{6}\;{\rm GeV}
~~~~&&
g^2={\footnotesize
\begin{pmatrix}
%(-0.324(2)-0.179(1)) \times 10^{-4}. & K^-p\\
%(-0.324(2)-0.179(1)) \times 10^{-4}&\bar{K}^0n\\
+0.118(15)\phantom{6}-i0.047(7)\phantom{65}&\pi^0\Lambda\\
+0.010(4)\phantom{65}+i0.008(3)\phantom{65}&\pi^+\Sigma^-\\
+0.010(4)\phantom{65}+i0.008(2)\phantom{65}&\pi^-\Sigma^+
\end{pmatrix}
}
\end{align}
%%%%%%%%%%%%%%
%%%%%%%%%%%%%%
with $1\sigma$ error bars from the re-sampling procedure. The $\Lambda$ pole positions compare well to those quoted in the PDG~\cite{ParticleDataGroup:2022pth}, particularly to those determined in chiral unitary models of the same type. For a discussion of chiral unitary model types see ~\cite{Mai:2020ltx}. Comparing the $\Lambda$ pole positions to the recent precision determination in Ref.~\cite{Lu:2022hwm}, the (narrower) $\Lambda(1405)$ poles agree, but there is only marginal overlap for the $\Lambda(1380)$, that is heavier and wider in the NNLO analysis of Ref.~\cite{Lu:2022hwm} than in the present analysis. It would be interesting to study the impact of the new data using that amplitude as well. 
The pole position of $\Sigma(1385)$ agrees well with the Breit-Wigner corrected determination~\cite{Lichtenberg:1974ja} quoted by the PDG~\cite{ParticleDataGroup:2022pth}
$(1379-1383)(1)-i(17-23)(2)~{\rm MeV}$.
The $g^2$ for the $\Sigma(1385)$ to the $\bar K N$ channel are not shown because they are of the order of $10^{-4}$ which is about two orders of magnitude smaller than the other couplings. We found that the reason lies in partial cancellations of terms in $\Gamma^{\bar KN}$ in Eq.~\eqref{eq:2pot}. The influence of the  $\Sigma(1385)$, being a P-wave, sub-threshold resonance is a-priori small, and its impact is further reduced by the tiny residue to $\bar KN$.

As discussed in our previous work~\cite{Sadasivan:2018jig} correlations between real and imaginary part of the pole positions can be substantial, such that any reasonable theoretical estimate should also provide corresponding correlation matrices. One reasonable way to provide such information is depicted in Fig.~\ref{fig:covcormat} for our central result $\mathfrak{F}_1$. 
The ellipses show the reduced confidence region when the correlation between the real and imaginary positions of each pole is accounted for. The colors in the top row visualise the relationship between poles. Each point of a given hue is from the same fit-sample visualizing cross-correlations between different poles. The fact that for the $\Lambda(1380)$ and  $\Lambda(1405)$, see top panel of Fig.~\ref{fig:covcormat}, there is a definite hue gradient in the points (e.g. blue points lie close to each other) indicates that the positions of the two poles are highly correlated. In contrast, the distribution of the points for the $\Sigma(1385)$-resonance does not have any noticeable pattern. This indicates that the correlation of the  positions of the $\Sigma(1385)$ with the positions of the $\Lambda(1380)$ and the $\Lambda(1405)$ poles is negligible.  
%%%%%%%%%%%%%%%%%%%
%%%%%%%%%%%%%%%%%%%

The covariance matrix, shown on the bottom panel of Fig.~\ref{fig:covcormat}, contains information about the precision of the fit parameters. The correlations are calculated from the covariance matrix as depicted also in the bottom panel of Fig.~\ref{fig:covcormat}. Both matrices use a heatmap visualization, i.e., the redder the pixel the more positive the correlation and the bluer the pixel the more negative the correlation. Indeed, we observe large off-diagonal elements only for   $\{{\rm Re}~W^*_{1380},{\rm Im}~W^*_{1380},{\rm Re}~W^*_{1405},{\rm Im}~W^*_{1405}\}$ which confirms the point distribution plots in the top row of the figure. The tilts of the ellipses show that the real and imaginary positions of the $\Lambda(1380)$ have a strong negative correlation and the real and imaginary positions of the $\Lambda(1405)$ show a weaker negative correlation. Furthermore, both the matrix and the coloring of the pole positions indicate that the real part of the $\Lambda(1380)$ is negatively correlated with both the real and imaginary part of the $\Lambda(1405)$. But the imaginary position of the $\Lambda(1380)$ is positively correlated with both components of the $\Lambda(1405)$ position.

%%%%%%%%%%%%%%%%%%%%%%%%%%%%%%%
\subsection{New data impact: poles and correlations}
\label{sec:data_vs_poles}
%%%%%%%%%%%%%%%%%%%%%%%%%%%%%%%
%%%%%%%%%%%%%%%%%%%%%%
%%%%%%%%%%%%%%%%%%%%%%
\begin{figure}[h!tbp]
\begin{center}
    \includegraphics[width=0.99\linewidth]{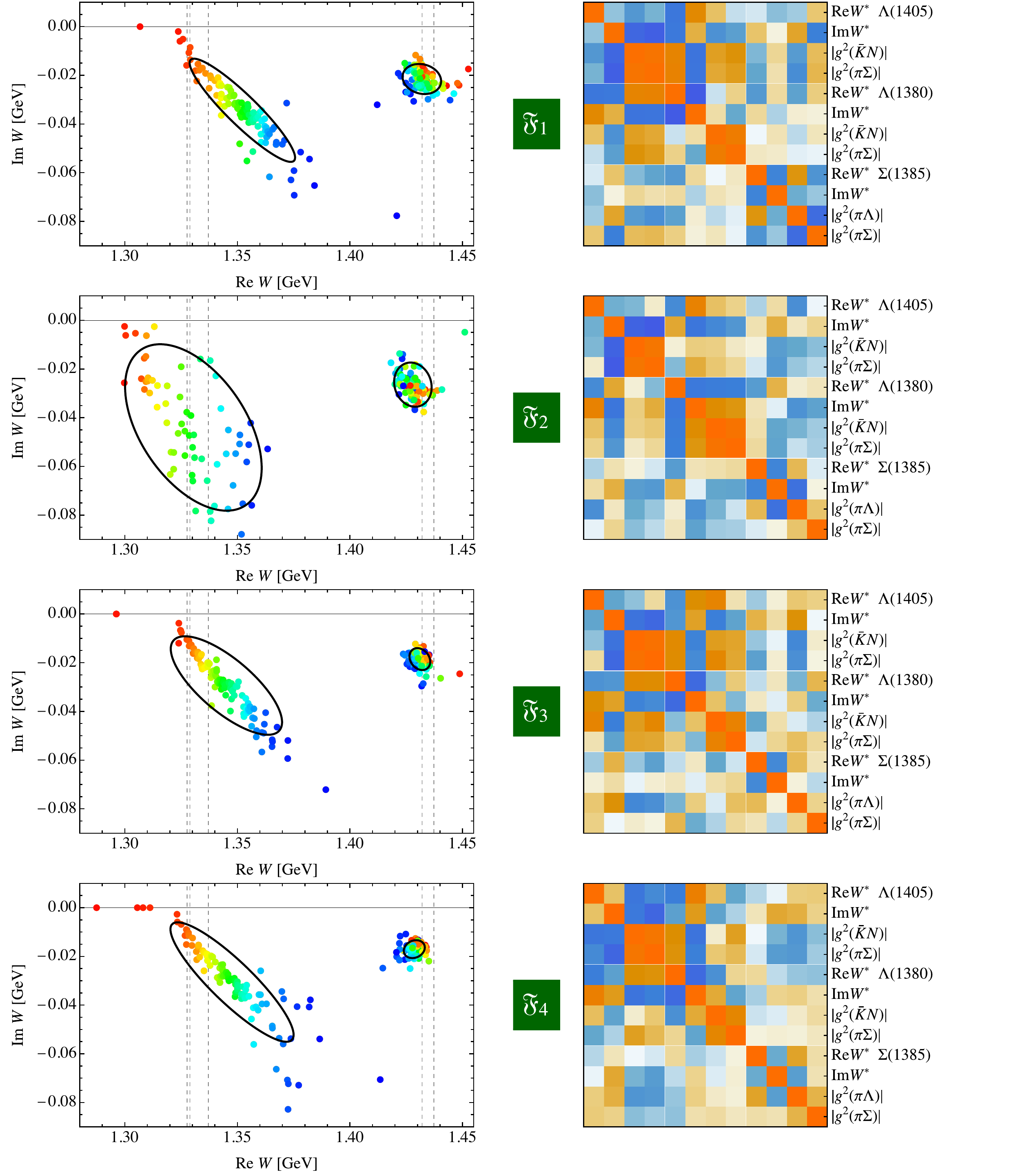}
    \caption{Comparison of all fit strategies. {\bf Left column}: Correlations between the pole positions of the $\Lambda(1380)$ and $\Lambda(1405)$ resonances for different fit strategies. Each point corresponds to a pole from a re-sampling solution. Points of the same hue belong to the same sample. {\bf Right column}: Correlations between pole positions and couplings. The redder the square the stronger the positive correlation and the bluer the square the stronger the negative correlation. }
    \label{fig:Bootstrap}
\end{center}
\end{figure}
%%%%%%%%%%%%%%%%%%%%%%
%%%%%%%%%%%%%%%%%%%%%%
Taking a step back, we turn now to a comparison of different fit strategies with respect to the
pole positions and couplings. In Fig.~\ref{fig:Bootstrap} we compare pole positions and correlations for the $I=0$ case $\{\Lambda(1380),\Lambda(1405)\}$. As in Fig.~\ref{fig:covcormat}, the ellipses give the uncertainty regions and the hues show the correlations. The correlations between the different cases are similar but not identical. One notable difference is that the error ellipse for the $\Lambda(1405)$ in case $\mathfrak{F}_4$ has a slight positive tilt whereas in the other cases, the $\Lambda(1405)$ ellipse has a slight negative tilt. The error ellipse for the $\Lambda(1380)$ in $\mathfrak{F}_2$ is significantly different from the ones of the other fits. Discrepancies for this fit from the others were already observed for the $K^-p\to K^-p$ amplitude in Fig.~\ref{fig:KbarNphys}.

The correlation matrix for each case is shown in the right column of Fig.~\ref{fig:Bootstrap}. It shows that the coupling for each pole to a given channel is very strongly correlated with all other couplings to that pole. The strongest correlation between  positions and couplings is the negative correlation between its imaginary position and the couplings to the $\Lambda(1405)$. This means that if the imaginary part of the  pole position moves further from the real axis, its residue is likely to increase.
In all four fit strategies there is a negligible correlation between the pole position (not considering the residues) of the $\Sigma(1385)$ and either isoscalar pole. In each fit strategy, there are 8 possible correlation coefficients between the pole positions of isovector to the isoscalar resonances for each of the four fits. These are all small enough that they could be explained by random variation in the bootstrap samples, even without any relationship between the positions of the resonances.
However, the residue of the $\Sigma(1385)$ to the $\pi \Lambda$ channel has stronger correlations with the $\Lambda(1405)$. In Fit $\mathfrak{F_1}$, the correlations to the imaginary pole position is 0.46, the correlation to the $\bar KN$ residue is $-0.25$, and the correlation to the $\pi\Sigma$ residue is $-0.33$. In fit $\mathfrak{F_3}$, these correlations respectively are $0.51$, $-0.52$, and $-0.47$, and in fit $\mathfrak{F_4}$, they are $0.32$, $-0.36$, and $-0.33$. These correlations are all statistically significant. On the other hand, there is no notable relationship between the $\Sigma(1385)$ and the $\Lambda(1380)$.   

The question arises whether or not it is necessary to include the $\Sigma(1385)$ in the fit, at all. To test this, we perform a refit starting with the parameters of $\mathfrak{F}_1$, called $\mathfrak{F}_1'$, in which we exclude the resonance by removing the pole term of Eq.~\eqref{eq:2pot} as well as the $\Sigma(1385)$ line-shape data~\cite{Baubillier:1984pj}. This results in the following values. There, the square bracket no longer indicate uncertainties but by how much the values in 
$\mathfrak{F}_1'$ changed compared to $\mathfrak{F}_1$ quoted in Eq.~\eqref{eq:F1-L1405-pole-coupling} and \eqref{eq:F1-L1380-pole-coupling}:
\begin{align}
&W^*_{\Lambda(1405)}=1.431[+1]\phantom{6}-i\,0.023[+0]\phantom{6}\;{\rm GeV}
~~~~&&
g^2={\footnotesize
\begin{pmatrix}
-0.103[-2]-i0.195[-2]&K^-p\\-0.092[-2]-i0.177[-6]&\bar{K}^0n\\+0.051[+3]+i0.038[-1]&\pi^0\Sigma^0\\+0.056[+1]+i0.036[+0]&\pi^+\Sigma^-\\+0.044[+3]+i\,0.040i[+0]&\pi^-\Sigma^+
\end{pmatrix}
}
\end{align}
%%%%%%%%%%%%%%
%%%%%%%%%%%%%%
\begin{align}
&W^*_{\Lambda(1380)}=1.357[+2]-i\,0.038[+0]\;{\rm GeV}
~~~~&&
g^2={\footnotesize
\begin{pmatrix}
-0.031[+7]+i0.144[+2] &K^-p\\
-0.028[+8]+i0.142[+2]&\bar{K}^0n\\
-0.110[+0]\phantom{6}+i0.103[+0]\phantom{6}&\pi^0\Sigma^0\\
-0.119[-1]\phantom{6}+i0.101[+1]\phantom{6}&\pi^+\Sigma^-\\
-0.103[-1]\phantom{6}+i0.100[-1]\phantom{6}&\pi^-\Sigma^+
\end{pmatrix}
}
\end{align}
%%%%%%%%%%%%%%
These values are very similar and well within the error bars of the previous values. This indicates that the inclusion of the $\Sigma(1385)$ has very limited impact on the $\Lambda(1380)$ and $\Lambda(1405)$ and it is not necessary to include the former in a determination of the latter.

The determinant of the covariance matrix, det~$C$, is the \emph{generalized variance} which is proportional to the square of the volume $V$ of the combined uncertainty region for fit parameters or extracted quantities~\cite{GeneralizedVariance},
%The full relationship for the $n$-dimensional volume $V^n$ for a given confidence region is 
\begin{align}
V=\frac{(K \pi)^{n/2}}{\Gamma\left(\frac{n}{2}+1\right)}\sqrt{\rm{det}~C}\,,
\end{align}
where $K$ is a constant related to the confidence level, $n$ is the dimension of the space,
$C$ is the covariance matrix and $\Gamma(x)$ Euler's gamma-function. This volume accounts for all possible correlations and is therefore a more accurate measurement than any uncertainty calculation that treats parameters independently. The size of this term is equivalent to the combined uncertainty in the values of the pole parameters which is given in the first row of Tab.~\ref{tab:generalized_variances}. 
This bulk measure confirms that the new data is useful for a precise determination of the pole parameters. 
%%%%%%%%%%%%%%%%%%%%%%%%
%%%%%%%%%%%%%%%%%%%%%%%%
\begin{table}[htb]
    \centering
    \scriptsize
    \begin{tabular}{llc}
    ~&$N_\text{pred.}$& $\mathfrak{F}_1$\\[1mm]  
    \toprule
   det~$C$ &18&  $2.12\times10^{-73}$   \\
    det~$C$$|_{\text{no correlations}}$ &18&  $1.62\times10^{-59}$  \\[2mm]

    det~$C$$|_{\text{pole positions}}$&4& $7.96\times10^{-19}$  \\
    det~$C$$|_{\text{pole positions}}|_{\text{no correlations}}$ &4& $2.56\times10^{-17}$ \\
    \bottomrule
    \end{tabular}
    \caption{The generalized variance, det~$C$, calculated from the covariance matrices for $\mathfrak{F_1}$. The second to fourth rows show generalized variances calculated for covariance matrices that have been reduced to isolate the effects of various correlations. Number of predicted resonance parameters is denoted $N_\text{pred.}$.}
    \label{tab:generalized_variances}
\end{table}
%%%%%%%%%%%%%%%%%%%%%%%%

%%%%%%%%%%%%%%%%%%%%%%%%

The 2nd to 4th rows of Tab.~\ref{tab:generalized_variances} give the generalized variances calculated
from the  reduced covariance matrices. For example, setting all off-diagonal values of the covariance matrix to 0 yields the values quoted in the 2nd row, i.e., neglecting all correlations. This isolates the effects of the correlations. The uncertainty region is reduced by a factor of  $\sqrt{(1.62 \times 10^{-59})/(2.12\times 10^{-73})}\approx9\times 10^{6}$. We emphasize that this very large increase in precision scales up with the number of predicted resonance parameters. Thus, comparing such quantities between models requires full knowledge of pole positions and couplings to different channels. Since this is hard to achieve in practice, we also determine the generalized variance considering only the positions of the $\Lambda(1380)$ and $\Lambda(1405)$ (real and imaginary part thereof) as predictions of the model. This is quoted in third row and neglecting correlations in the fourth row of  Tab.~\ref{tab:generalized_variances}. In case $\mathfrak{F}_1$ the correlations result in a decrease in the uncertainty region by a factor of $\sqrt{(2.56 \times 10^{-17})/(7.96\times 10^{-19})}\approx6$. In summary, a reasonable comparison of the uncertainties between different models requires one to determine correlations between predicted resonance parameters.

The generalized variance can also be used to compare how constrained the different fits are. The full generalized variances for all fit scenarios read
\begin{align}
    &\text{det}~C_{\mathfrak{F}_1}=2.12\times 10^{-73}\,,~~ 
    \text{det}~C_{\mathfrak{F}_2}=1.36\times 10^{-59}\,,~~\\
    &\text{det}~C_{\mathfrak{F}_3}=2.52\times 10^{-71}\,,~~
    \text{det}~C_{\mathfrak{F}_4}=3.24\times 10^{-70}\,.
\end{align}
We note that fit $\mathfrak{F_2}$ has a much larger generalized variance than the others. This larger uncertainty region can be seen from the pole positions in Fig.~\ref{fig:Bootstrap}. With the other three fits, the addition of  data points constrains the uncertainty region. Including both new data sets reduces the uncertainty region by a factor of 
$\sqrt{\text{det}~C_{\mathfrak{F}_4}}/\sqrt{\text{det}~C_{\mathfrak{F}_1}}\approx 39$.

The $\chi^2_a$ in $\mathfrak{F}_4$ for the new $\sigma_{K^-p\to\pi^0 \Lambda}$ and $|f_{0+}^{\pi^-\Lambda \to K^-n}|$ data are 15.20 and 5.08, respectively ($\mathfrak{F}_4$ is not fit to these points). In $\mathfrak{F}_3$, these values are reduced to 0.89 and 1.76, however, this reduction in $\chi^2_a$ involves an increase in the total $\chi^2_{\rm dof}$ showing a bit of tension. Fit scenario $\mathfrak{F}_2$ has the best $\chi^2_{\rm dof}\approx 1.06$, however, this fit is also not ideal. The generalized variance of this fit, shown in Tab.~\ref{tab:generalized_variances}, is 16 orders of magnitude larger than the other fits and the generalized variance for only pole positions of the $\Lambda(1380)$ and $\Lambda(1405)$ is two orders of magnitude larger. This much larger uncertainty region can be seen in the larger ellipses and weaker correlations in Fig.~\ref{fig:Bootstrap}. The large uncertainty for $\mathfrak{F}_2$ could be due to over-fitting the point $|f_{0+}^{\pi^-\Lambda \to K^-n}|$ which has a partial $\chi^2_a$ of only 0.02, reduced from 5.63 in $\mathfrak{F}_4$. Fit $\mathfrak{F}_1$ which considers both new sets of data has a better $\chi^2_{\rm dof}\approx1.19$ and a smaller generalized variance. This supports the importances of including both new data sets~\cite{Piscicchia:2018rez,Piscicchia:2022wmd}.
In conjunction, the two new data sets allow for a fit that simultaneously describes all (old and new) data and more tightly constrains the pole positions of the $\Lambda(1380)$, $\Lambda(1405)$ and $\Sigma(1385)$.

 \subsection{Belle $\pi^{\pm}\Lambda$ line-shape data}
 \label{sec:belle}
 The Belle collaboration recently measured $\Lambda\pi^+$ and $\Lambda\pi^-$ line-shapes from $\Lambda_c^+\to \Lambda\pi^+\pi^+\pi^-$ decays~\cite{Belle:2022kvv}. For the first time, narrow structures at the $\bar K N$ thresholds are resolved that appear as small enhancements on top of the right shoulder of the large $\Sigma(1385)$ resonance. These isospin $I=1$ structures appear at slightly different masses for the two line-shapes, potentially reflecting the different thresholds coming from mass differences within kaon and nucleon multiplets, respectively. Note that such isospin-1 structures were first considered in Ref.~\cite{Oller:2000fj}
 in the context of chiral-unitary approaches.

 %While many $S$-wave transitions in our amplitude show cusp structures at the $\bar KN$ thresholds the question is whether 
 
 %%%%%%%%%%%%%%%%%%%%%%
 \begin{figure}[htb]
     \begin{center}
     \includegraphics[width=0.6\linewidth]{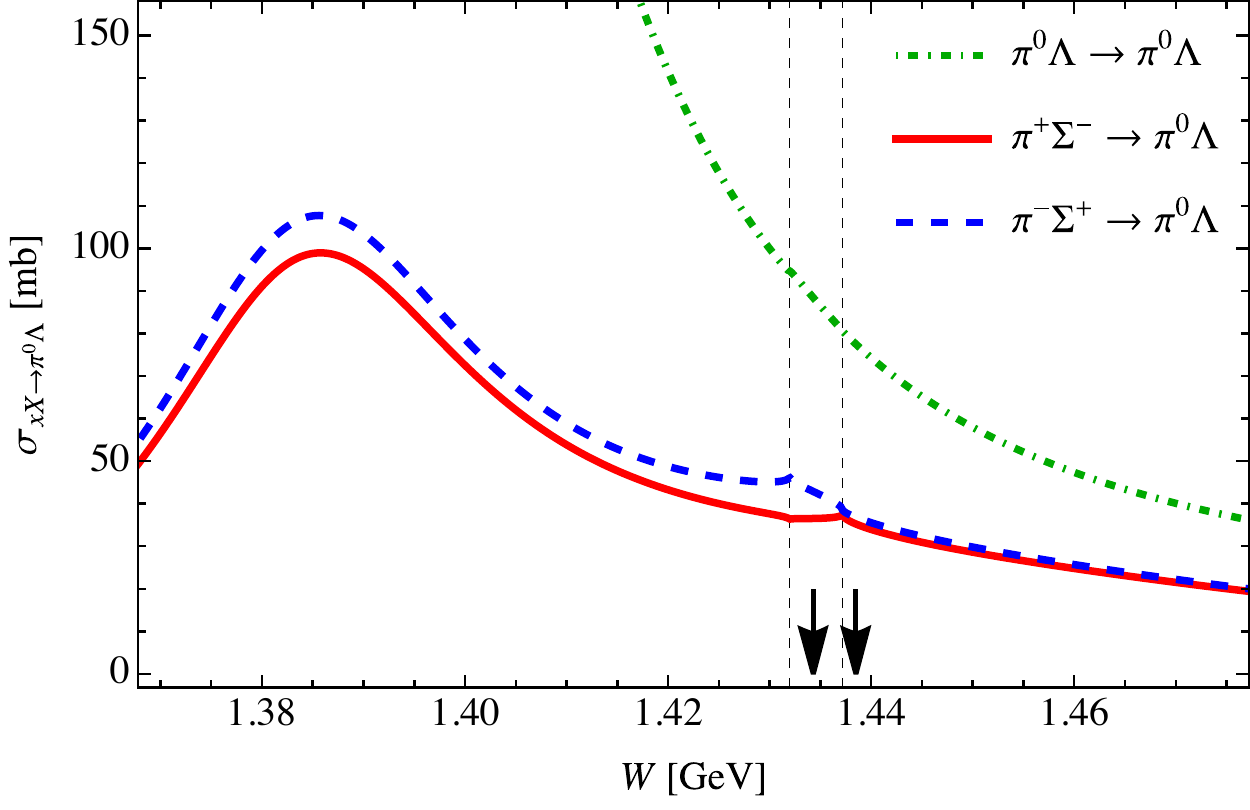}
     \end{center}
     \vspace*{-0.5cm}
 \caption{Cross sections (containing S- and P-waves) for different initial states to the $\pi^0\Lambda$ final state with respect to fit scenario $\mathfrak{F}_1$ (all data fit). Dashed vertical lines show positions of $K^-p$ and $\bar K^0n$ thresholds, while the black arrows indicate the positions of the possible isovector resonances from Ref.~\cite{Belle:2022kvv}.}
 \label{fig:belleline}
 \end{figure}
%%%%%%%%%%%%%%%%%%%%%%
 The present amplitude explicitly contains the $\Sigma(1385)$ with realistic mass and width fitted to  $\pi\Lambda$ line-shape data~\cite{Baubillier:1984pj} as shown in Fig.~\ref{fig:Amplitudes}. Note that in that picture we do not show our total cross section but only the squared P-wave amplitude (i.e., no S-wave with potential cusps). In fact, our isospin-1 amplitudes exhibit S-wave threshold cusps in $\pi\Lambda$. Some transitions incuding both S-wave and P-wave are shown in Fig.~\ref{fig:belleline}.
For the $\pi\Sigma\to\pi^0\Lambda$ transitions we observe a similar pattern as in the new Belle data~\cite{Belle:2022kvv}, i.e., small cusp structures with peaks at slightly different masses, on top of the large shoulder of the $\Sigma(1385)$. For the $\pi^0\Lambda\to\pi^0\Lambda$ transition, the $\Sigma(1385)$ is so dominant that the S-wave cusps disappear entirely, as the figure shows.

We leave the comparison at this qualitative level, because a more quantitative analysis requires to formulate our amplitude for total charges $Q=\pm 1$ while here we have it only available at $Q=0$. In addition, the actual data will be described by a superposition of the processes shown in the figure, including even other ones not shown, such as $K^-n\to\pi^-\Lambda$ and $\bar K^0p\to\pi^+\Lambda$. This involves new fit parameters, similarly as needed in the description of other line-shape data~\cite{Hemingway:1984pz}. We leave this to future work.

%%%%%%%%%%%%%%%%%%%%%%%%%%%%%%%
\section{Conclusion}
\label{sec:summ}
%%%%%%%%%%%%%%%%%%%%%%%%%%%%%%%

We analyse the impact of new data from the KLOE and AMADEUS experiments. We also include other, previously not used data, putting stringent constraints on the line-shape of $\Sigma(1385)$. Using these data we re-fit the chiral NLO unitary coupled-channel model.
The new pole positions for the $\Lambda(1380)$, $\Lambda(1405)$, and $\Sigma(1385)$ are consistent with the pole positions of previous analyses quoted by the current edition of the particle data group review. 

The impact of each set of new data is studied in detail in fit scenarios that exclude them, showing their effect on the poles of the $\Lambda(1380)$, $\Lambda(1405)$, and the $\Sigma(1385)$. 
The new data do not further constrain the pole positions of the two $\Lambda$ states much. However, the overall uncertainty of pole parameters (including residues), as encoded in the generalized variance, is reduced by a factor of 40. This would be equivalent to a reduction of the uncertainty of pole parameters by 20\%  on average by the new KLOE and AMADEUS data.

As for the $\Sigma(1385)$, there are some correlations of its coupling with some parameters of the two $\Lambda$ states. However, that does not mean that this resonance must be included in the analysis. Indeed, by omitting it and the associated line-shape data, the pole parameters of the $\Lambda$ states change only well within uncertainties (pole positions by less than 2~MeV).

For the first time, we determine correlations between resonance parameters, in particular of the two $\Lambda(1405)$ states. These correlations are as important as error bars, and we show that for a proper comparison of different models it is necessary to include them. Indeed, the generalized uncertainty $\sqrt{\text{det}C}$ decreases by a factor of six if correlations of pole positions are taken into account.

In addition, we made an initial comparison with recently measured Belle line-shape data for the $\pi^\pm\Lambda$ final states. We observe cusp structures at the $\bar KN$ thresholds on top of the right shoulder of the $\Sigma(1385)$. Future work with the amplitude formulated in non-zero net charge will allow for quantitative studies. 

Comparing our pole position of the $\Lambda(1380)$ with the NNLO results of Ref.~\cite{Lu:2022hwm} we observed some tension. It would be interesting to update that work using the new data from KLOE and AMADEUS which would also allow for a better determination of the systematics of chiral unitary approaches. 

%%%%%%%%%%%%%%%%%%%%%%%%%%%%
%%%%%%%%%%%%%%%%%%%%%%%%%%%%
\section*{Funding}
This work of MM and UGM was supported by Deutsche Forschungsgemeinschaft (DFG, German Research Foundation), the NSFC through the funds provided to the Sino-German Collaborative Research Center CRC 110 “Symmetries and the Emergence of Structure in QCD” (DFG Project-ID 196253076 - TRR 110, NSFC Grant No.~12070131001). The work of MD and MM is supported by the National Science Foundation under Grant No. PHY-2012289. The work of MD is also supported by the U.S. Department of
Energy grant DE-SC0016582
and DOE Office of Science, Office of Nuclear Physics
under contract DE-AC05-06OR23177. LSG is partly supported by the National Natural Science Foundation of China under Grant No.11735003, No.11975041,  and No. 11961141004. JXL acknowledges support from the National Natural Science Foundation of China under Grant No.12105006 and China Postdoctoral Science Foundation under Grant No. 2021M690008. 
UGM was further supported by funds from the
CAS through a President’s International Fellowship Initiative (PIFI) (Grant No. 2018DM0034) and by the VolkswagenStiftung (Grant No. 93562)

\section*{Acknowledgments}
Special thanks to Anthony Gerg for helping with this research.
%%%%%%%%%%%%%%%%%%%%%%%%%%%%
%%%%%%%%%%%%%%%%%%%%%%%%%%%%
\clearpage

\appendix
\section*{Appendix: Further results}
%%%%%%%%%%%%%%%%%%%%%%
%%%%%%%%%%%%%%%%%%%%%%
\begin{table*}[htbp]
\centering
\renewcommand{\arraystretch}{0.5}
\begin{tabular}{|l|c|c|c|c|}
\hline
Parameter & $\mathfrak{F}_1$ (all new data) &  $\mathfrak{F}_2$ (new amp) &  $\mathfrak{F}_3$ (new cs) &  $\mathfrak{F}_4$ (no new data)\TT\BB \\
\hline
$a_1$ & $+0.2543$& $+0.7631$  & $+0.1916$ & $+0.2910$ \TT\\
$a_2$ & $+1.1890$ & $+1.6767$  & $+1.1655$ & $+1.6867$ \\
$a_3$ & $+0.1158$ & $-0.2048$  & $+0.1674$ & $+0.1612$ \\
$a_4$ & $-1.5820$ & $-0.3661$  & $-0.8428$ & $-0.7976$ \\
$a_5$ &$-2.5870$ & $-3.2600$  & $-2.2306$ &  $-1.7103$ \\
$a_6$ &$-1.3710$ & $-1.7491$  & $-1.3372$ & $-1.2320$ \\
$b_1$ & $-0.3813$ & $-0.3567$  & $-0.3753$ & $-0.3554$ \\
$b_2$ & $+1.3979$  & $+0.6008$ & $+1.3035$ & $+0.9854$\\
$b_3$ & $-0.2878$ & $-0.2332$  & $-0.2958$ & $-0.2828$ \\
$b_4$ &$-0.3012$ & $-0.2162$  & $-0.2798$ & $-0.2476$ \\
$b_5$ &$+0.2502$ & $+0.1186$ &  $+0.2910$ & $+0.2514$ \\
$b_6$ &$-0.7228$ & $-0.4414$ & $-0.8125$ & $-0.7952$ \\
$b_7$ &$-0.8582$ & $-0.4403$ & $-1.0135$ &  $-1.0332$ \\
$b_8$ & $+0.0436$  & $+0.0447$ & $+0.0508$ & $+0.0857$ \\
$b_9$ &$-0.4800$ & $-0.2879$ & $-0.5488$ & $-0.4557$ \\
$b_{10}$ & $-0.0314$ & $+0.0094$ & $-0.0283$ &   $+0.0044$\\
$b_{11}$ & $+0.4471$ & $+0.104$8 & $+0.4591$ & $+0.3926$ \\
$b_0$ &$-0.8892$  &$-1.0699$  & $-0.8799$ & $-0.8550$ \\
$b_D$ & $+0.2775$& $+0.4028$ & $+0.2714$ & $+0.3562$ \\
$b_F$ &$-0.0001$ & $-0.3473$ & $-0.1323$ & $-0.2446$ \\
$\lambda\,[{\rm GeV}^{-1}]$ &$+2.2982$ & $+2.5776$ & $+2.4062$  &
$+3.3518$ \\
$m_\Sigma^0\,[{\rm GeV}]$ & $+1.5660$ & $+1.6592$ & $+1.5857$
& $+1.8961$\\
$f_\Sigma$ & $+0.6378$ & $+0.5285$ & $+0.5362$ & $+0.3975$\BB\\
\hline
\end{tabular}
\caption{Best fit parameters for the four fit scenarios. These parameters are accessible in digital form in auxiliary arXiv-files.}
\label{tab:parameters}
\end{table*}
%%%%%%%%%%%%%%%%%%%%%%
%%%%%%%%%%%%%%%%%%%%%%

%%%%%%%%%%%%%%%%%%%%%%
%%%%%%%%%%%%%%%%%%%%%%
\begin{table*}[htbp]
\scriptsize
% \hspace{0cm}
\addtolength{\tabcolsep}{-0.5mm}
\renewcommand{\arraystretch}{1.3}
\centering
\begin{tabular}{|l|l|r|r|r|r|}
\hline
& & $\mathfrak{F}_1$ (all new data) &  $\mathfrak{F}_2$ (new amp) &  $\mathfrak{F}_3$ (new cs) &  $\mathfrak{F}_4$ (no new data) \\
\hline
\cline{1-6}
$\Lambda(1380)$&$g_{K^- p}^2  $&$ -0.038+0.146i $&$-0.373+1.407i $&$-0.052+0.1133i  $&$-0.076+0.107i$\TT\\
&$g_{\bar{K}^0 n}^2$ &$ -0.036+0.144i $&$ -0.375+1.480i $&$ -0.052+0.116i  $&$ -0.076+0.111i
$\\
&$g_{\pi^0 \Sigma^0}^2$&$ -0.110+0.103i $&$ -0.098+0.480i $&$ -0.101+0.104i $&$   -0.096+0.094i
$\\
&$g_{\pi^+ \Sigma^-}^2$&$ -0.118-0.102i $&$ -0.106+0.469i $&$ -0.111+0.103i  $&$ -0.108+0.092i
$\\
&$g_{\pi^- \Sigma^+}^2$&$ -0.102+0.101i $&$ -0.093+0.478i $&$ -0.091+0.102i $&$   -0.085+0.092i
$\\
\hline
$\Lambda(1405)$&$g_{K^- p}^2$&$ -0.101-0.193i $&$ -0.040-0.283i $&$  -0.097-0.133i $&$ -0.085-0.103i
$\\
&$g_{\bar{K}^0 n}^2$&$ -0.090-0.171i $&$ -0.025-0.243i $&$ -0.081-0.124i  $&$ -0.066-0.095i
$\\
&$g_{\pi^0 \Sigma^0}^2$ &$ +0.048+0.039i $&$ +0.077+0.007i $&$  +0.031+0.036i $&$ +0.027+0.032i
$\\
&$g_{\pi^+ \Sigma^-}^2$ &$ +0.055+0.036i $&$ +0.081-0.001i $&$ +0.034+0.035i  $&$ +0.030+0.030i
$\\
&$g_{\pi^- \Sigma^+}^2$ &$ +0.041+0.040i $&$ +0.072+0.015i $&$ +0.027+0.038i  $&$ +0.023+0.033i
$\\
\hline
$\Sigma(1385)$&$g_{\pi^0\Lambda^0}^2$ &$ +0.118-0.047i $&$ +0.130-0.047i $&$ +0.129-0.053i  $&$ +0.147-0.006i
$\\
&$g_{\pi^+ \Sigma^-}^2$&$ +0.010-0.008i $&$ +0.003-0.003i $&$ +0.007-0.005i  $&$ +0.004-0.003i
$\\
&$g_{\pi^- \Sigma^+}^2$&$ +0.010-0.008i $&$ +0.003-0.003i $&$  +0.007-0.005i $&$ +0.004-0.003i$
\\
\hline
\end{tabular}
\caption{
\label{tab:chi2-allx}
Individual and total $\chi^2$ for the fit strategy $\mathfrak{F}_1,\dots, \mathfrak{F}_4$. The individual contributions to the $\chi^2$ are the $\chi^2_a$ which contributes to the $\chi^2_\text{dof}$ as in Eq.~\eqref{eq:chi2un}. Predicted observables (not minimized $\chi_a^2$ contributions) are put in parentheses. Bottom part of the table collects the predicted pole positions $W^*\in\mathds{C}$ and the pole residues $g^2$. Uncertainties on the pole parameters in $\mathfrak{F}_1$ can be determined from the covariance matrix given in the auxiliary arXiv-files as described in Sec.~\ref{sec:pole-correlations}.
}
\end{table*}
%%%%%%%%%%%%%%%%%%%%%%
%%%%%%%%%%%%%%%%%%%%%%

%%%%%%%%%%%%%%%%%%%%%%
%%%%%%%%%%%%%%%%%%%%%%
\begin{figure}[htb]
    \includegraphics[width=0.45\linewidth]{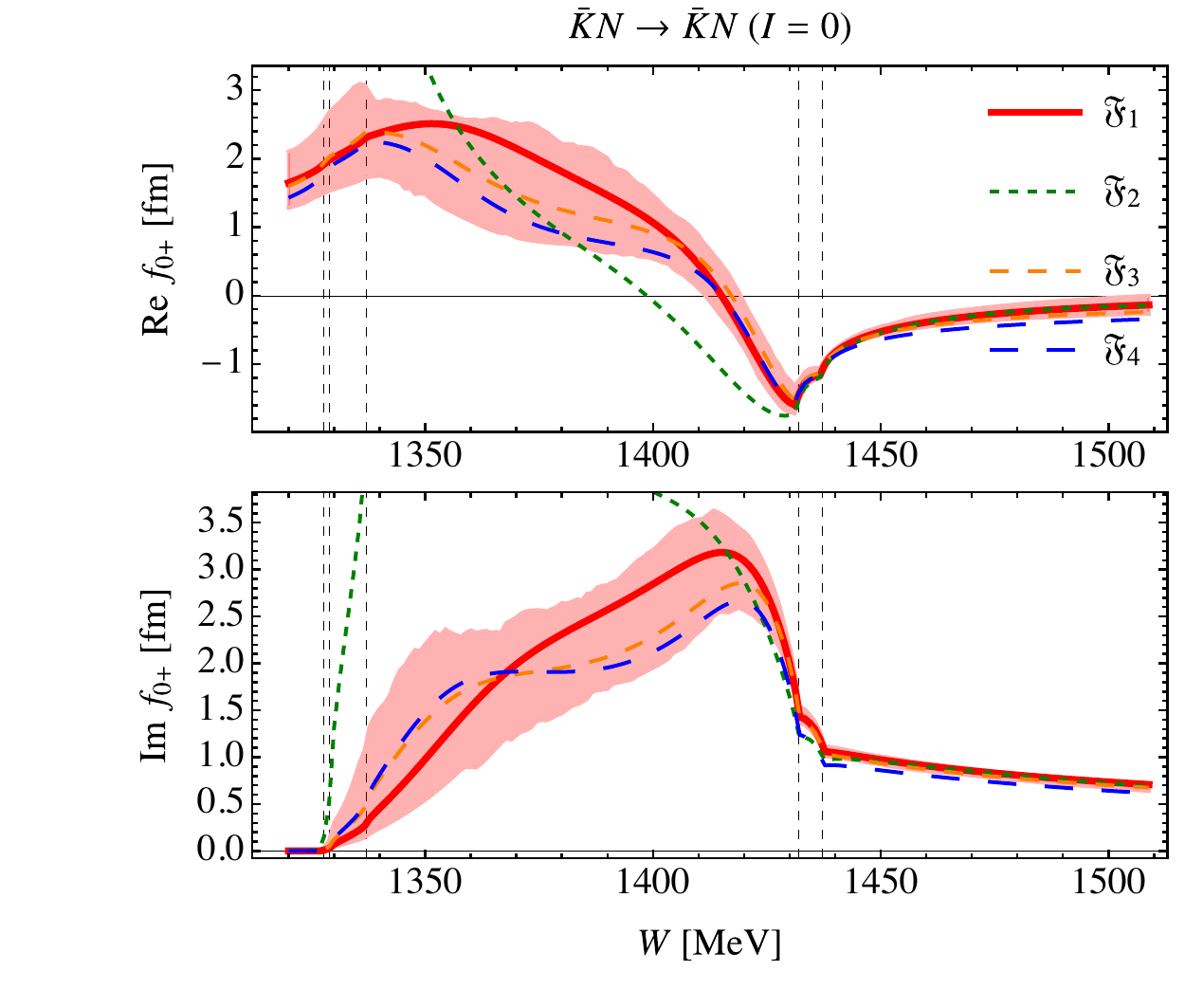}
    \includegraphics[width=0.45\linewidth]{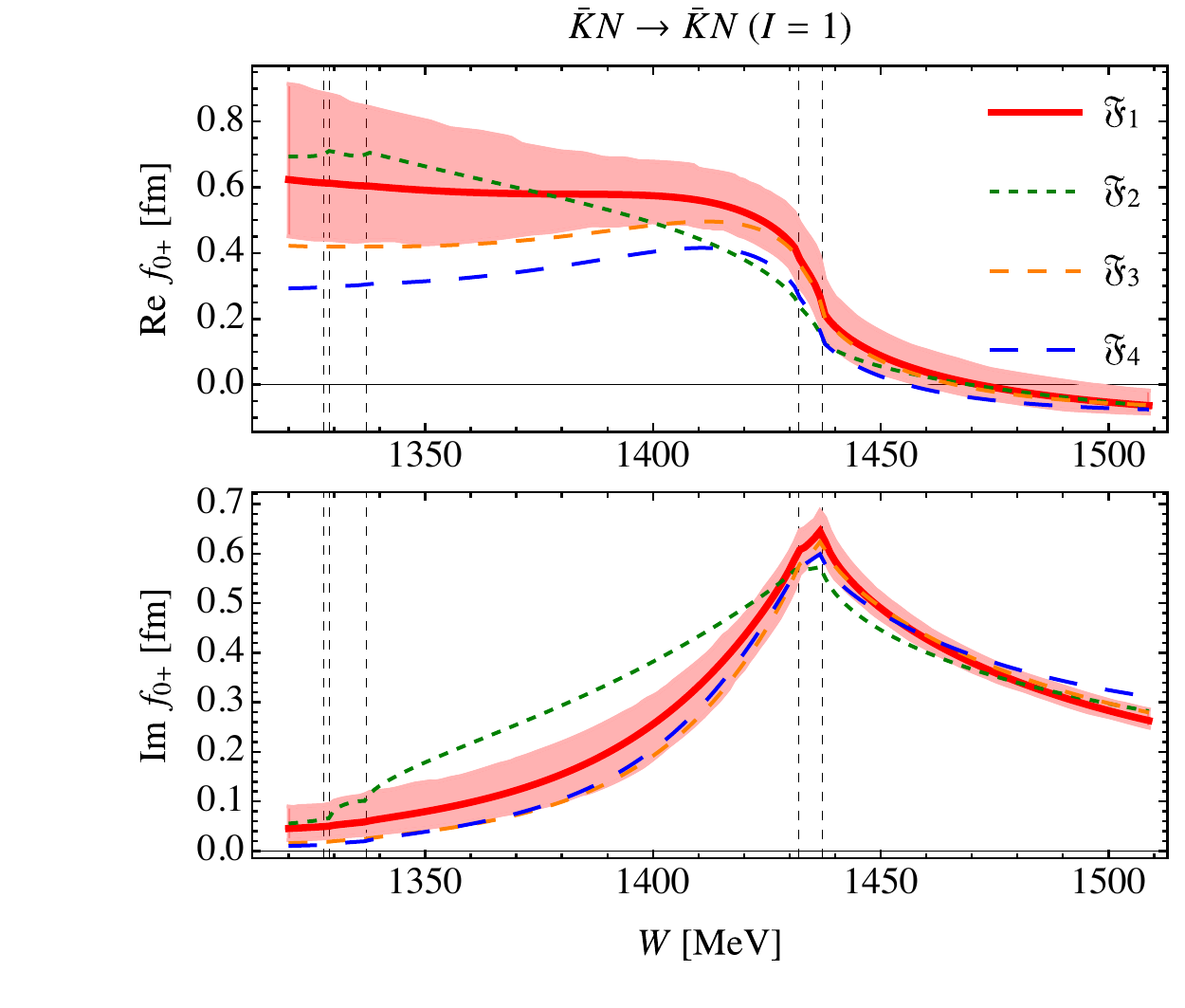}
    \includegraphics[width=0.45\linewidth]{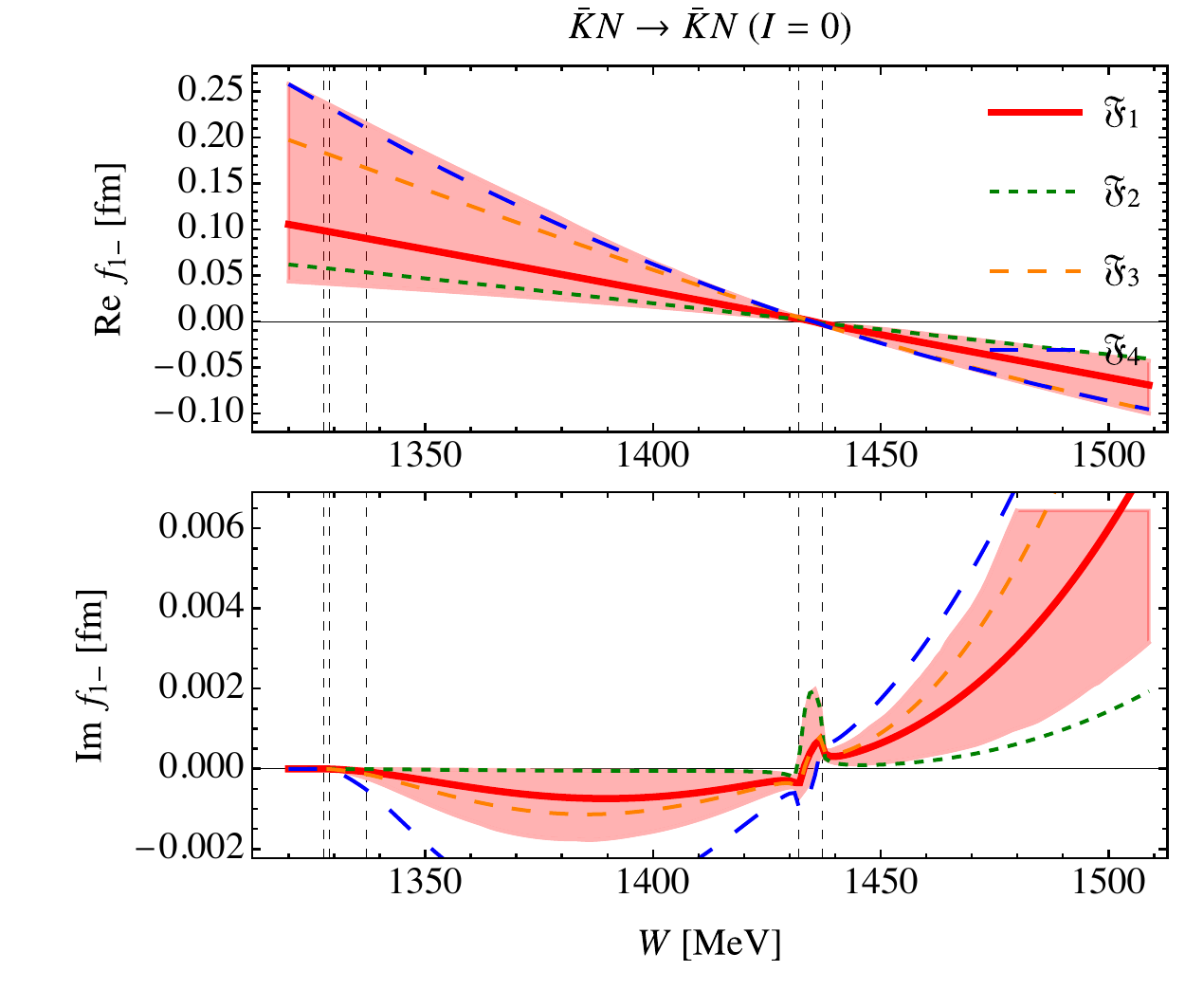}
    \includegraphics[width=0.45\linewidth]{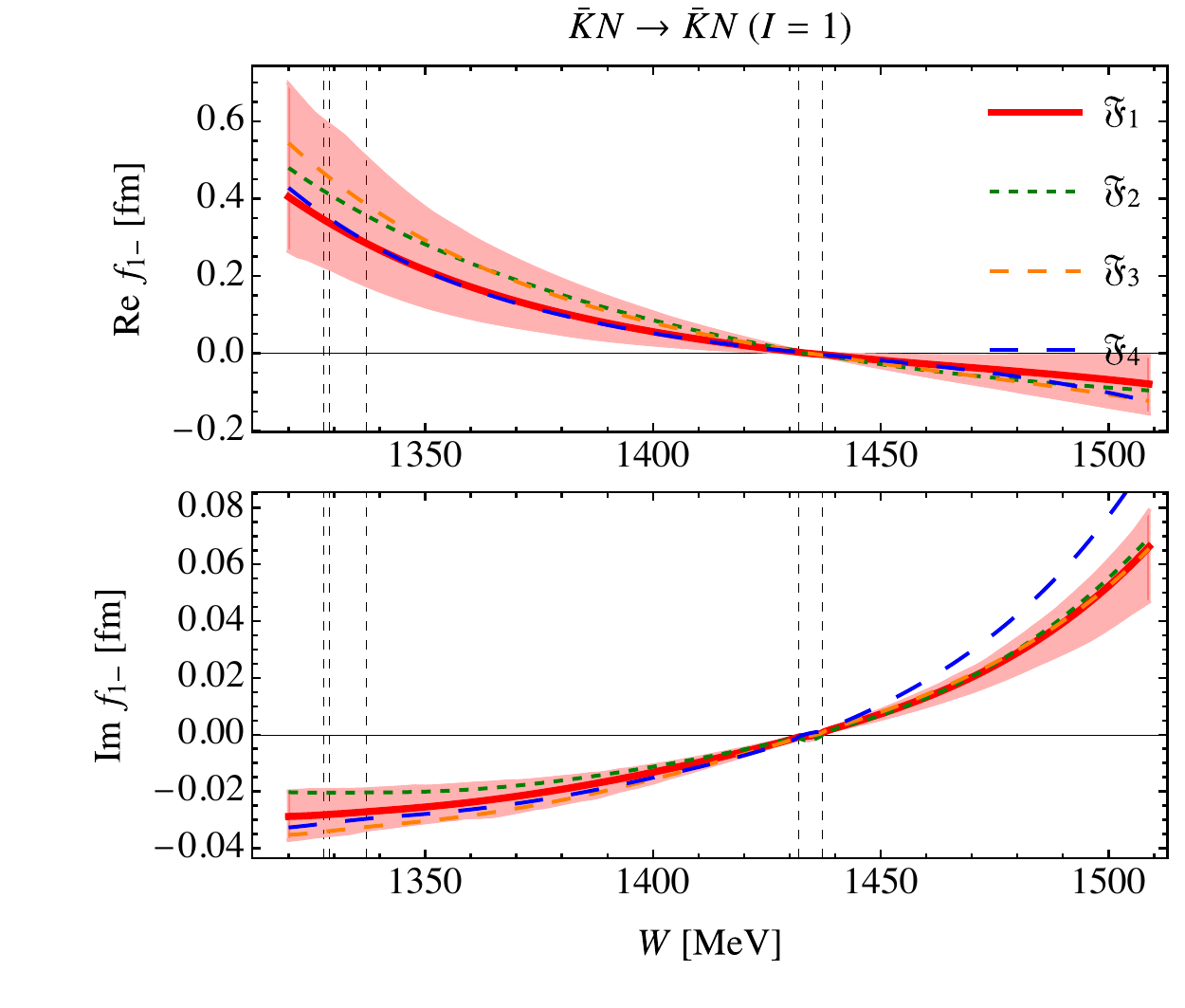}
    \includegraphics[width=0.45\linewidth]{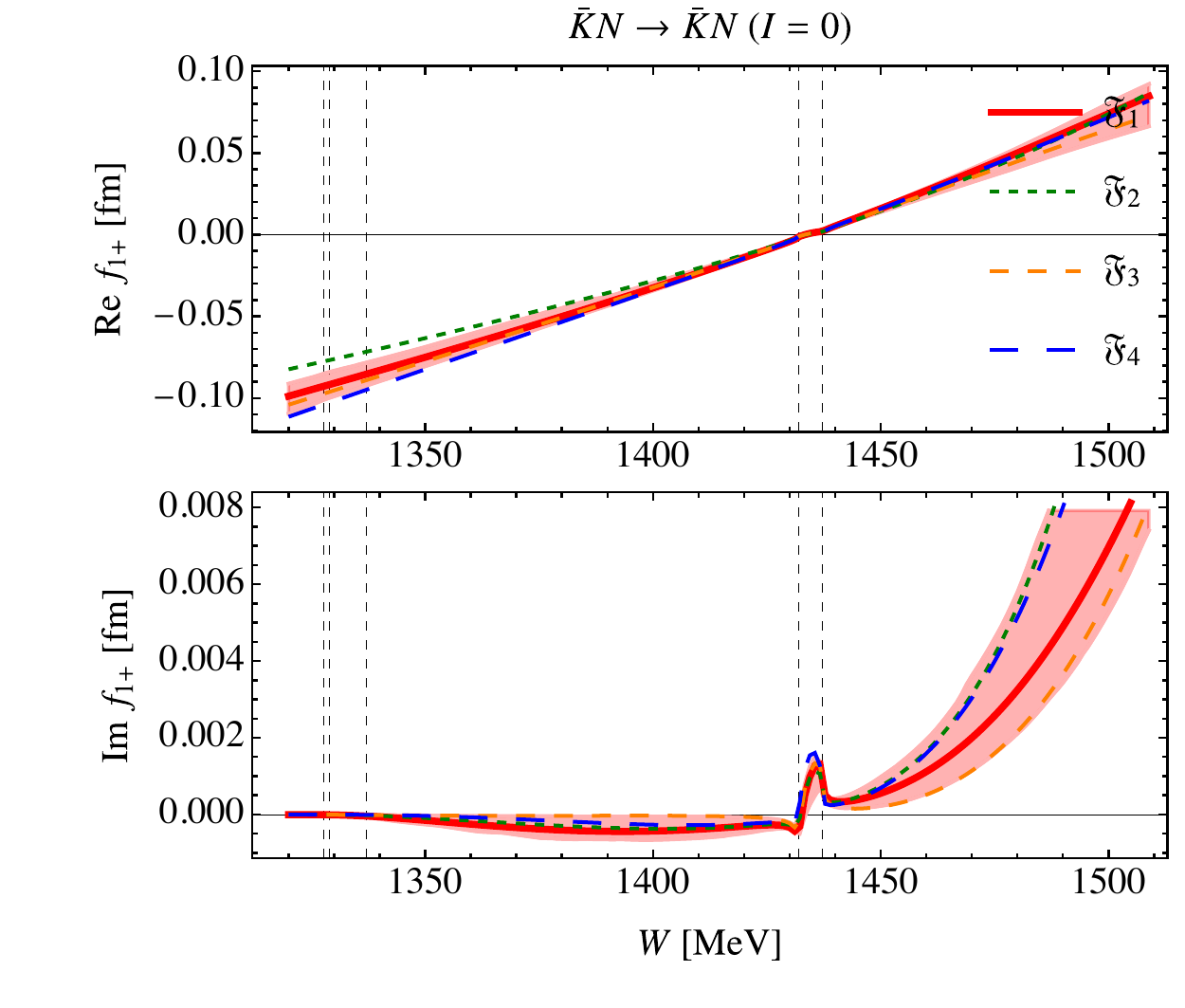}\hspace{+1.6cm}
    \includegraphics[width=0.45\linewidth]{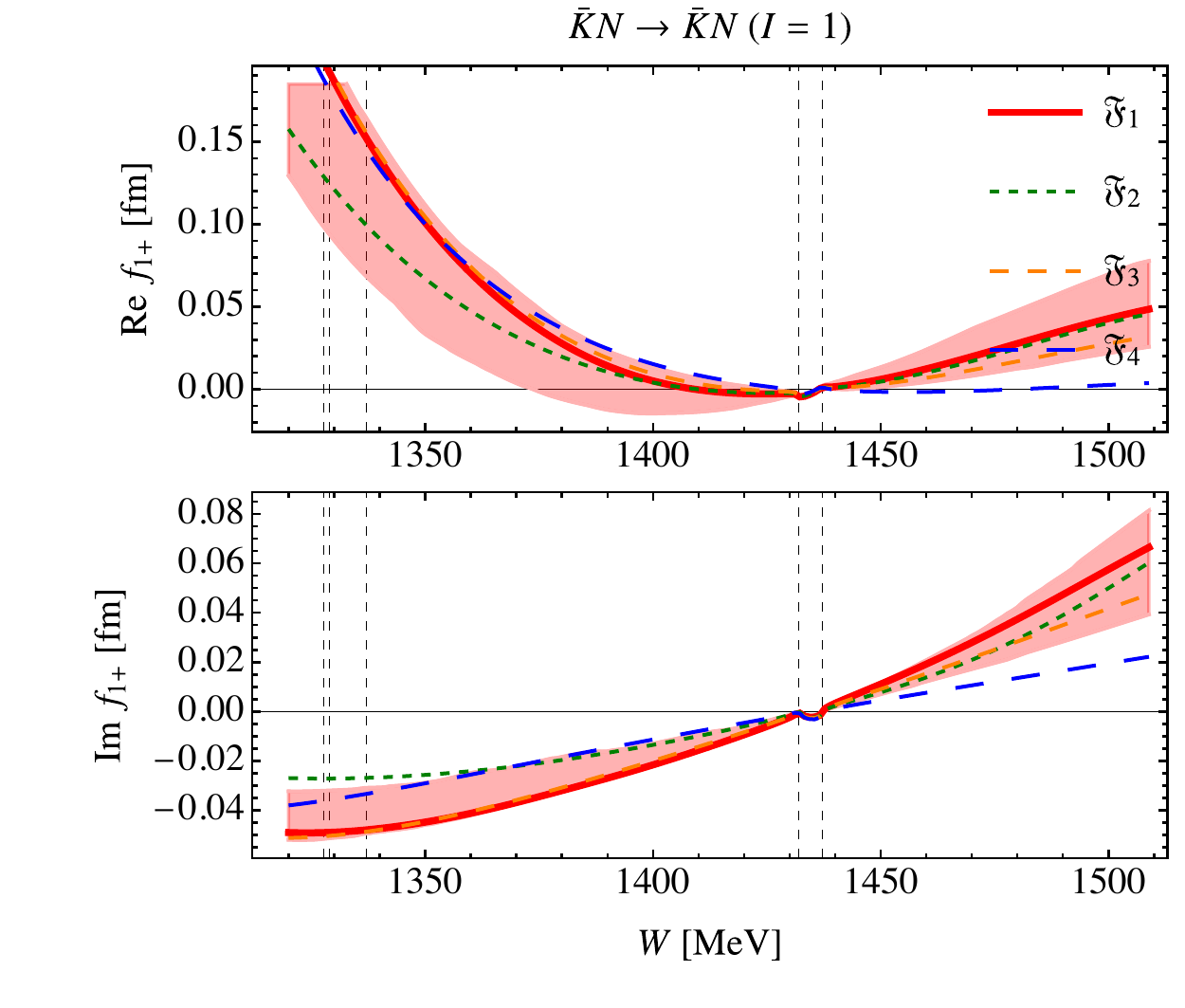}
    \caption{Isospin channels of the S- and P-wave amplitudes in the four fit scenarios.}
    \label{fig:KbarNIsospin}
\end{figure}
%%%%%%%%%%%%%%%%%%%%%%
%%%%%%%%%%%%%%%%%%%%%%
\clearpage

%%%%%%%%%%%%%%%%%%%%%%%%%%%%
%%%%%%%%%%%%%%%%%%%%%%%%%%%%
% \bibliographystyle{Frontiers-Vancouver}
\bibliographystyle{JHEP}
\bibliography{bib.bib}
%%%%%%%%%%%%%%%%%%%%%%%%%%%%
%%%%%%%%%%%%%%%%%%%%%%%%%%%%

\end{document}